\documentclass[
    prx, 
    aps, reprint,
    twocolumn,groupedaddress,superscriptaddress,
    footinbib,
    floatfix,
    longbibliography,
    showpacs,showkeys,
    ]{revtex4-2}

\usepackage{amssymb,amsmath,graphicx}
\usepackage{dcolumn}% Align table columns on decimal point
\usepackage{bm}% bold math
\usepackage[tight]{subfigure}
\usepackage{xcolor}

\usepackage[bookmarksnumbered,
colorlinks = true,
linkcolor = black,
urlcolor  = blue,
citecolor = blue,
anchorcolor = blue]{hyperref}

\newcommand{\HS}{\hat{H}_\mathrm{S}}
\newcommand{\HSB}{\hat{H}_\mathrm{SB}}
\newcommand{\HB}{\hat{H}_\mathrm{B}}
\newcommand{\Ho}{\hat{H}}
\newcommand{\Uo}{\hat{U}}
\newcommand{\Heff}{\hat{H}_\mathrm{eff}}
\newcommand{\rme}{\mathrm{e}}
\newcommand{\rmi}{\mathrm{i}}
\newcommand{\diffd}{\mathrm{d}}
\newcommand{\llangle}{\langle\!\langle}
\newcommand{\rrangle}{\rangle\!\rangle}

\makeatletter
\def\maketitle{
    \@author@finish
    \title@column\titleblock@produce
    \suppressfloats[t]}
    \makeatother

\makeatletter
\begin{document}

\title{Dissipative Floquet engineering of gapped many-body phases using thermal baths}
\author{Lorenz Wanckel}
\email{wanckel@tu-berlin.de}
\author{André Eckardt}
\email{eckardt@tu-berlin.de}
\affiliation{Technische Universität Berlin, Institut für Physik und Astronomie, 10623 Berlin, Germany}

\begin{abstract}
    Floquet engineering, the control of a quantum system by means of time-periodic driving, allows to modify the properties of a system so that it is described by an approximate effective time-independent Hamiltonian. 
    However, in the presence of interactions the stabilization of interesting many-body ground states of such effective Hamiltonians is possible only on a certain time scale, beyond which Floquet heating sets in, as it results from unwanted driving induced resonant excitation. Moreover, the preparation of the ground state of such Floquet engineered effective Hamiltonians usually has to be accomplished adiabatically, which is another source of heating, especially when a phase transition has to be passed in a large system.
    Here, we propose a general dissipative strategy for the preparation and stabilization of effective ground states that are protected by an energy gap, such as topologically ordered states. 
    It is scalable and relies on coupling the driven system to a thermal bath, the properties of which are chosen so that it both suppresses Floquet heating and guides the system into a non-equilibrium steady state with a large occupation of the effective ground-state. The small residual occupations of effective excited states are non-thermal, reflecting the non-equilibrium nature of the scheme. 
    We use the Floquet-Born-Markov master equation to verify the proposed strategy. In particular, we discuss the preparation of both a one-dimensional Mott insulator and a two-dimensional fractional Chern insulator in strongly driven Bose-Hubbard systems.
\end{abstract}

\date{\today}
\maketitle
\hypersetup{linkcolor=blue}

\section{Introduction}
When applying a rapid time-periodic drive to a many-body quantum system \cite{GoldmanDalibard2014, bukovUniversalHighfrequencyBehavior2015,eckardtHighfrequencyApproximationPeriodically2015, eckardtColloquiumAtomicQuantum2017, OkaKitamura2019, WeitenbergSimonet2021}, we can, roughly speaking, expect three effects. First of all, the system will follow the drive, giving rise to fast ``micromotion'' at the driving frequency. Second, on slower time scales the system properties are modified in a way that results from averaging over this micromotion. 
This effect can be captured by an approximate effective time-independent Hamiltonian $\Ho_\text{eff}$, usually obtained from a low-order high-frequency expansion. 
Controlling quantum systems in this way is known as Floquet engineering. Beyond simple time averaging, it can be described using systematic high-frequency expansions~\cite{GoldmanDalibard2014,bukovUniversalHighfrequencyBehavior2015,eckardtHighfrequencyApproximationPeriodically2015,MikamiEtAl2016}.

The third effect is that the drive induces resonant transitions, corresponding to the absorption or emission of driving quanta $\hbar\Omega$, with angular driving frequency $\Omega$. For $\hbar\Omega$ large compared to the typical single- and two-particle energy scales of the Hamiltonian, such processes require a substantial change of the system's wave function (corresponding to a high-order process in perturbation theory), so that the corresponding time scale can become large.
Nevertheless, in generic many-body systems these resonant processes make themselves felt eventually. This is  known as Floquet heating \cite{bukovUniversalHighfrequencyBehavior2015, eckardtHighfrequencyApproximationPeriodically2015, eckardtColloquiumAtomicQuantum2017,OkaKitamura2019,WeidingerKnapp2017,ReitterEtAl2017,SinghEtAl2019,WintersbergerEtAl2020,IkedaPokovnikov2021,RakcheevLaeuchli2022}. Even though the associated heating time increases exponentially with increasing driving frequency (i.e.\ with increasing order of perturbation theory required to describe excitations at energy $\hbar\Omega$) \cite{abaninExponentiallySlowHeating2015}, it cannot become arbitrarily large. Namely, the frequency also needs to remain small compared to large energy scales associated with particles leaving the system, such as the energy of high-lying orbital states neglected in a low-energy tight-binding description of a lattice system \cite{WeinbergEtAl2015,StraeterEckardt2016}. This competition of heating effects for too low and too high frequencies, respectively, leads to an optimal window of driving frequencies \cite{sunOptimalFrequencyWindow2020}, where the heating time takes its maximum value.  In addition to choosing favorable frequencies, Floquet heating can also be attenuated by exploiting the coherent suppression of dominant heating channels \cite{eckardtAvoidedLevelCrossingSpectroscopyDressed2008}, destructive interference from two-tone driving \cite{ViebahnEtAl2021}, or the suppression of multi-photon transitions via spectral filtering \cite{Shimasaki2024}. However, in the thermodynamic limit, generic isolated Floquet systems are expected to approach an infinite-temperature state in the long-time limit, as a consequence of eigenstate thermalization without energy conservation \cite{LazaridesEtAl2014, DAlessioRigol2014}.

\begin{figure*}
    \begin{center}
    \includegraphics[width=0.95\linewidth]{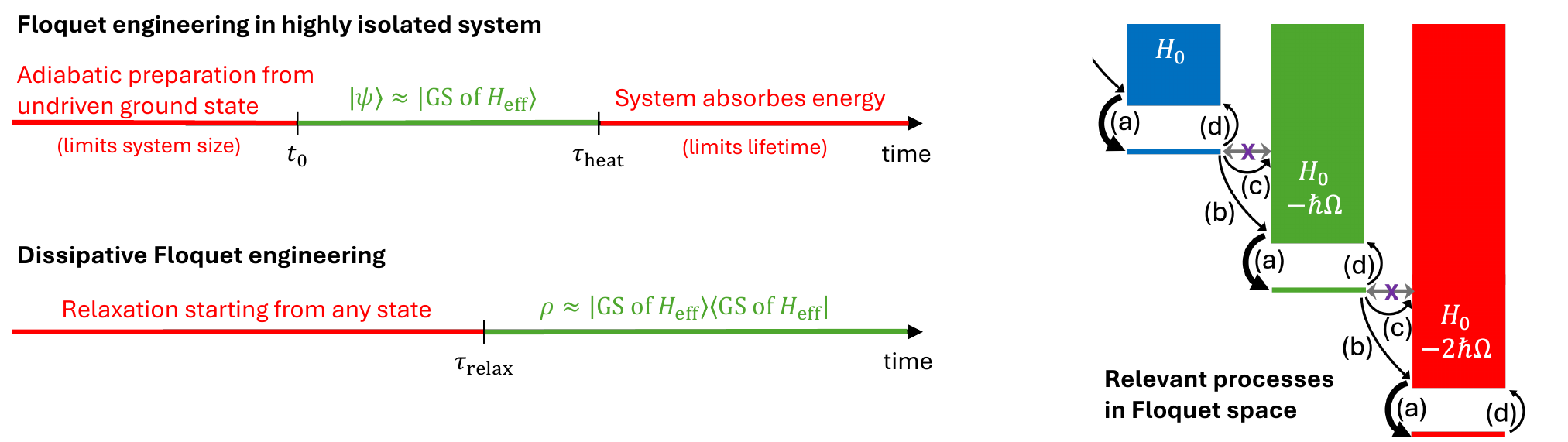}
    \caption{
    \label{fig:TwoProtocols}
    \label{fig:FloquetSpace}
 Left:~Preparation of the ground state (GS) of $\Ho_\text{eff}$ both via traditional Floquet engineering in isolated systems, limited by adiabatic preparation and Floquet heating, and via dissipative Floquet engineering of the ground state. 
 Right:~Sketch of the most relevant processes in Floquet space. The different colors indicate the spectra of $\Ho_0+m\hbar\Omega$ for different ``photon'' numbers $m$. For each copy the ground state is separated by an energy gap $E_\text{gap}$ from excited states. The gray solid straight double arrows describe the most relevant resonant coherent coupling of the ground states at $m$ to excited states at $m'=m-1$. This is the source of Floquet heating. The purple crosses indicate the suppression of these processes via the coupling to the bath. The curved black single arrows symbolize the most relevant incoherent bath-induced transitions, with (b-d) engineered to be suppressed relative to (a), leading to a large ground-state occupation $P_0$.}
    \end{center}
\end{figure*}

In recent years, high-frequency Floquet engineering has been employed successfully in different experimental platforms, such as ultracold atoms in optical lattices or photons in waveguides and superconducting circuits. It was used, among others, for the control of the bosonic Mott transition \cite{eckardtSuperfluidInsulatorTransitionPeriodically2005, zenesiniCoherentControlDressed2009}, studying frustrated tunnel kinetics in a triangular lattice \cite{EckardtEtAl2010,StruckEtAl2011}, the realization of artificial magnetic fields and topological band structures \cite{aidelsburgerExperimentalRealizationStrong2011,
aidelsburgerRealizationHofstadterHamiltonian2013,
jotzuExperimentalRealizationTopological2014, 
aidelsburgerMeasuringChernNumber2015,
taiMicroscopyInteractingHarper2017, 
tarnowskiMeasuringTopologyDynamics2019}, the control of localization \cite{Shimasaki2024,DottiEtAl2025},
the realization of the anyon Hubbard model for two particles \cite{kwanRealizationOnedimensionalAnyons2024,BakkaliHassaniEtAl2026}, as well as the preparation of ``baby'' fractional Chern insulators of up to three particles \cite{GrushinEtAl2014, leonardRealizationFractionalQuantum2023,wangRealizationFractionalQuantum2024,kwanPfaffianQuantumHall2026}. Recently, also the interactions between ultracold atoms were modified using the Floquet control of Feshbach resonances \cite{GuthmannEtAl2025}. However, the stabilization of correlated states of matter of more than just a few particles over a longer time remains elusive. Apart from Floquet heating, here a second problem is the need to adiabatically prepare the ground state of $\Ho_\text{eff}$, starting from the ground state of the undriven system. Especially, when a phase transition is passed, heating can often only be suppressed with the help of finite-size energy gaps in rather small systems. Moreover, during adiabatic protocols it is not always possible to avoid the passing parameter regimes, where resonant Floquet heating is prominent~\cite{eckardtAvoidedLevelCrossingSpectroscopyDressed2008}.

A possible strategy for the preparation and stabilization of interesting Floquet engineered states of matter is the controlled use of dissipation. For cooling a non-driven system close to its ground state, it can be coupled to a cold thermal bath or a suitably engineered driven-dissipative environment (like in the case of laser cooling). In both cases, a situation is achieved in which it is relatively easy for the environment to absorb energy from the system, whereas it is more difficult to emit energy, unless the system has already reached the desired low-energy state. However, these strategies cannot simply be adapted in a straightforward fashion to Floquet engineered systems, since here Floquet heating occurs also at the level of system and environment together. Namely, via the absorption of driving quanta $\hbar\Omega$, ``easy'' bath-induced transitions become possible that are prohibited by energy conservation in undriven systems, which increase the energy in the bath but at the same time also in the system. As a consequence of these processes, open driven systems generically approach steady states that are neither approximately thermal (with respect to the approximate effective Hamiltonian) nor low-energy (see, e.g., Refs.~\cite{breuerQuasistationaryDistributionsDissipative2000,KetzmerickWustmann2010,  vorbergGeneralizedBoseEinsteinCondensation2013,schnellFloquetheatinginducedBoseCondensation2023,MatsyshynEtAl2023}). 

For small Floquet systems, with an upper energy bound well below $\hbar\Omega$, these unwanted transitions can, however, be suppressed by choosing also an environment with a spectral width that is narrow enough, so that not even together system and bath can absorb a driving quantum $\hbar\Omega$. This strategy also works for non-interacting particles, as long as their single-particle energies are bounded. It has been proposed for the stabilization of Floquet engineered topological insulators of free fermions \cite{iadecolaOccupationTopologicalFloquet2015, seetharamControlledPopulationFloquetBloch2015, qinChargeDensityWave2018, schnellDissipativePreparationFloquet2024}.  
It fails, however, for large, interacting systems. Yet, it has been shown in Ref.~\cite{shiraiEffectiveFloquetGibbs2016} for the example of a spin chain that the resonant absorption of energy can still be suppressed by a thermal bath also in larger systems, leading to an approximate thermal state of the effective Hamiltonian $\Ho_\mathrm{eff}$, as long as the system-bath coupling remains small compared to the energy level splitting of $\Ho_\mathrm{eff}$ (see also Ref.~\cite{moriFloquetStatesOpen2023}). In this approach the coupling to the environment suppresses resonant excitation within the system \cite{honeStatisticalMechanicsFloquet2009}, while further conditions ensure that also the environment cannot absorb driving quanta $\hbar\Omega$. However, as we explain below, this strategy becomes difficult for the relevant case of gapped effective ground states in the interesting regime where the bath temperature $\beta^{-1}$ becomes small compared to the energy gap $E_\text{gap}$.

In this paper, we present a general and simple strategy for approximately preparing and stabilizing gapped ground states of effective Hamiltonians of interacting many-body Floquet systems via the coupling to a thermal bath. This includes the relevant case of topologically ordered states of matter, such as fractional Chern insulator phases in systems with both strong Floquet engineered artificial magnetic fields and strong interactions. The differences between such dissipative Floquet engineering and traditional Floquet engineering in highly-isolated systems are sketched on the left-hand side of Fig.~\ref{fig:TwoProtocols}. 
For the dissipative protocol, we require the excited states of $H_\mathrm{eff}$ to be occupied weakly, but not to approximately follow a quasi-thermal state $\propto \exp(-\beta H_\mathrm{eff})$. This turns out to be possible under realistic conditions in a steady state that is stabilized by a steady energy flow from the driven system into the bath. Notably, our approach does only require the system-bath coupling to be small compared to $E_\text{gap}$, but not relative to the level splitting between excited states, making it scalable to large system sizes. We discuss and verify it for the simple example of a driven Bose-Hubbard chain at unit filling. The effective Hamiltonian of this model possesses a gapped Mott-insulator ground state, when interactions are strong relative to the Floquet controlled effective tunneling matrix element \cite{eckardtSuperfluidInsulatorTransitionPeriodically2005}. Moreover, we also demonstrate the dissipative preparation of a Floquet engineered bosonic fractional Chern-insulator state of interacting bosons in a Floquet engineered quarter-flux Harper-Hofstadter model ~\cite{leonardRealizationFractionalQuantum2023}.

\section{Floquet engineering}
Consider a quantum system described by a time-periodic Hamiltonian 
\begin{equation}
\Ho(t)=\Ho(t+T)=\sum_{\mu=-\infty}^\infty \rme^{\rmi\mu\Omega t} \Ho_\mu, 
\end{equation}
with driving period $T$, angular frequency $\Omega=2\pi/T$, and Fourier components $\Ho_\mu=\frac{1}{T}\int_0^T\!\diffd t\,\Ho(t)\rme^{-\rmi\mu\Omega t}$ with integer $\mu$. It possesses quasi-steady solutions (Floquet states) of the form $|\psi_n(t)\rangle=|u_n(t)\rangle\rme^{-\rmi \varepsilon_n t/\hbar}$, with time-periodic Floquet modes $|u_{n}(t)\rangle=|u_{n}(t+T)\rangle$ and real quasienergies $\varepsilon_n$ \cite{shirleySolutionSchrodingerEquation1965}. These can formally be obtained by finding a generalized gauge transformation $\Ho_F=\Uo_F^\dag\Ho\Uo_F-\rmi\hbar\Uo_F^\dag\diffd_t\Uo_F$ with a time-periodic unitary micromotion operator $\Uo_F(t)=\Uo_F(t+T)$, so that the transformed Hamiltonian, $\Ho_F$, becomes time independent. From the eigenvalue problem $\Ho_F|\tilde{u}_n\rangle = \varepsilon_n|\tilde{u}_n\rangle$, one then gets both Floquet modes $|u_n(t)\rangle=\Uo_F(t)|\tilde{u}_n\rangle$ and quasienergies $\varepsilon_n$ (see, e.g.\ Ref.~\cite{eckardtColloquiumAtomicQuantum2017}).

The quasienergies are defined modulo $\hbar\Omega$ only, so that each Floquet state generates an infinite ladder of equivalent modes $|u_{nm}(t)\rangle=|u_{n0}(t)\rangle\rme^{\rmi m\Omega t}$ with quasienergies $\varepsilon_{nm}=\varepsilon_{n0}+m\hbar\Omega$. Plugging these solutions into the time-dependent Schrödinger equation, we see that they form independent solutions of the eigenvalue problem $\hat{Q}|u_{nm}\rrangle=\varepsilon_{nm}|u_{nm}\rrangle$, where the quasienergy operator $\hat{Q}=\HS(t)-\rmi\hbar \diffd_t$ 
acts in the extended Hilbert space of time-periodic states \cite{sambeSteadyStatesQuasienergies1973}, with scalar product $\llangle\cdot|\cdot\rrangle=\frac{1}{T}\int_0^T \diffd t \langle\cdot|\cdot\rangle$. We denote elements of this Floquet (or Sambe) space by double kets, $|\cdot\rrangle$. Using basis states $|k m\rrangle$, corresponding to $|k\rangle \rme^{\rmi m\Omega t}$, where $\{|k\rangle\}$ is a generic basis of the original state space, one finds matrix elements
\begin{equation}\label{eq:Q}
\llangle k'm'|\hat{Q}|k m\rrangle 
= \langle k'|\Ho_{m'-m}|k\rangle +\delta_{m'm}\delta_{k'k} m\hbar\Omega,
\end{equation}
which feature a transparent block structure with respect to the index $m$. The diagonal blocks correspond to the time-averaged Hamiltonian $\hat{H}_0$ shifted by $m\hbar\Omega$, resembling the problem of a system with Hamiltonian $\Ho_0$ coupled to a photon-like mode with energy $m\hbar\Omega$ and $m$ playing the role of a photon number (relative to a large background occupation). The individual spectra of different diagonal blocks are also sketched on the right-hand side of Fig.~\ref{fig:FloquetSpace}. 
The off-diagonal blocks described by $\Ho_{\mu\ne0}$ correspond to $\mu$-``photon'' processes. For a time-independent Hamiltonian, with $\Ho_{\mu\ne0}=0$, $\hat{Q}$ is block diagonal. For non-vanishing $\Ho_{\mu\ne0}$, it can be block-diagonalized by a unitary rotation, which is equivalent to the above  transformation $\Uo_F$ 
\cite{eckardtHighfrequencyApproximationPeriodically2015}.

For $\hbar\Omega$ large compared to both the spectral width of $\Ho_0$ and the coupling given by the matrix elements of $\Ho_{\mu\ne0}$, we can systematically block diagonalize $\hat{Q}$ by using van Vleck degenerate perturbation theory in Floquet space. Choosing the ``photon'' energy $m\hbar\Omega$ as the unperturbed problem, this corresponds to a high-frequency expansion \cite{eckardtHighfrequencyApproximationPeriodically2015}. For simplicity, we will consider the leading non-trivial order for both $\Ho_F$ and $\Uo_F$ \footnote{Here the approximation of $\Ho_F$ corresponds to a lower order than that for $\Uo_F$. This is fine. As long as further corrections to $\Ho_F$ remain small and do not describe qualitatively new effects, we can approximately neglect them.}:
\begin{align}\label{eq:HF}
    \Ho_F\simeq \Ho_0\equiv \Ho_\text{eff}, \quad \Uo_F\simeq \rme^{-\sum_{\mu\neq 0} \frac{\rme^{\rmi \mu\Omega t}}{\mu\hbar\Omega} \Ho_\mu}\equiv \Uo_F^{\mathrm{eff}}. 
\end{align}
Considering a higher-order approximation is technically more involved, but not conceptually \footnote{To include also high-frequency corrections in $\Ho_\text{eff}$ beyond the leading-order given by $\Ho_0$, one can perform a unitary rotation of $\hat{Q}$ with a finite-order approximation of $\Uo_F^{\mathrm{eff}}$, leading to diagonal blocks given by $\Ho_0$ plus these corrections. Subsequently, one can continue with the analysis, as it is discussed here.}.
We call the approximate time-independent Hamiltonian effective Hamiltonian $\Ho_\mathrm{eff}$. Its eigenvalue problem, $\Ho_\text{eff}|\tilde{v}_\kappa\rangle = E_\kappa |\tilde{v}_\kappa\rangle$, defines effective eigenstates and energies, from which we obtain approximate Floquet modes $|v_\kappa(t)\rangle = \Uo_F^{\mathrm{eff}}(t)|\tilde{v}_\kappa\rangle$ with approximate quasienergies $E_\kappa$. The effective energies $E_\kappa$ correspond to the time averaged energies of these approximate Floquet states, their increase, thus, indicates heating.

In an interacting many-body system with excited states at extensive energies, we cannot expect $\hbar\Omega$ to be large compared to the spectral width of $\Ho_0$. The approximation $\Ho_F\approx\Ho_\text{eff}$ remains, however, useful, as long as $\hbar\Omega$ is still large compared to the typical energy change $\mathcal{E}_{\Ho_\mu}$ (with respect to the time-averaged Hamiltonian $\Ho_0$), as it is associated with transitions induced by applying $\Ho_{\mu}$ to relevant states. For definiteness, we might define 
\begin{align}\label{eq:EnergyScale}
\mathcal{E}_{\hat{X}}\equiv\big|\langle\tilde{v}_0'|\Ho_0|\tilde{v}_0'\rangle-\langle\tilde{v}_0|\Ho_0|\tilde{v}_0\rangle\big|,
\quad
|\tilde{v}_0'\rangle \equiv \frac{\hat{X}|\tilde{v}_0\rangle}{\lVert\hat{X}|\tilde{v}_0\rangle\rVert}
\end{align}
for any operator $\hat{X}$, with the relevant state being the ground state of $\Ho_\text{eff}$, which is our target state. For a many-body system with two-particle interactions, applying $\Ho_\mu$ to a many-body wave function will change the state of one or two particles, so that $\mathcal{E}_{\Ho_\mu}$ remains intensive. Thus, as long as 
\begin{align}\label{eq:ConditionFlouqetEngineering}
\mathcal{E}_{\Ho_\mu}\ll\hbar\Omega,
\end{align}
the resonant $\mu$-``photon'' coupling (with $\mu\ne 0$) between two eigenstates of $\Ho_\text{eff}$ that are separated in energy by (about) $\mu\hbar\Omega$ will be associated with tiny matrix elements of $\Ho_{\pm\mu}$ only. 

For $\mu=\pm1$, this is illustrated on the right-hand side of Fig.~\ref{fig:FloquetSpace}, where we depict the spectra of $\Ho_0+m\hbar\Omega$ with different $m$ by different colors. The ground state of $\Ho_0$ at photon number $m$ (blue and green for  $m=0$ and $m=-1$, respectively) is degenerate with an excited state of $\Ho_0$ of energy $\hbar\Omega$ at photon number $m'=m-1$ (green and red for $m'=-1$ and $m'=-2$, respectively), to which it is coupled via $\Ho_{-1}$,  while the hermitian conjugated reverse process is described by $\Ho_{1}=\Ho_{-1}^\dag$. This coupling, indicated by grey double arrows, is small as long as $\mathcal{E}_{\Ho_1}\ll\hbar\Omega$. The corresponding transitions, which describe resonant Floquet heating, will therefore occur only after a relatively long time.

\begin{figure}
    \begin{center}
    \includegraphics[width=0.95\linewidth]{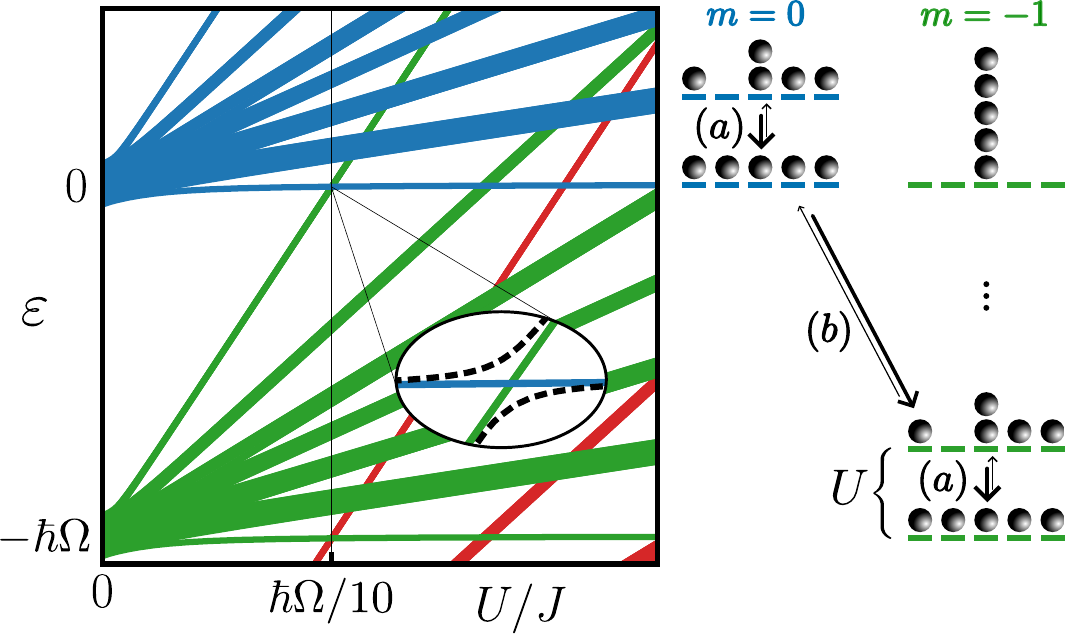}
    \caption{\label{fig:SpectrumSketch}
    Sketch of a many-body quasienergy spectrum at high frequencies, using the example of the driven Bose-Hubbard model, with scaled interaction strength $U/J$. In a first approximation it is given by the overlapping spectra of $\Ho_\text{eff}+m\hbar\Omega$ (shown with different colors for different $m$). For large $U/J$ it consists of bands above a gapped ground state, corresponding to different numbers and configurations of particle-hole excitations (as those sketched on the right) dispersed by tunneling. Resonant coupling leads to tiny avoided crossings between states of different $m$ (inset). The dominant bath-induced transitions into and out of the $m=0$ copy of the $\Ho_\text{eff}$ ground state are labeled $(a)$ and $(b)$, respectively.} 
    \end{center}
\end{figure}

According to these considerations, when depicting the spectrum of $\hat{Q}$ as a function of a parameter in this high-frequency regime, it can be understood as given by infinitely many copies of the spectrum of $\Ho_\text{eff}$ shifted by $m\hbar\Omega$ (Fig.~\ref{fig:SpectrumSketch}), with tiny avoided crossings between states of different ``photon'' number $m$. In  Fig.~\ref{fig:SpectrumSketch} this is sketched for the case of a five-site driven Bose-Hubbard chain at unit filling, where the inset indicates an avoided crossing between the effective ground state with a band of effective excited states characterized by with five particles on the same lattice site.

The approximate description of a driven quantum system by the effective Hamiltonian $\Ho_\text{eff}$ obtained from low-order high-frequency expansion provides an important strategy for Floquet engineering that has been used successfully in quantum simulators of atoms and photons, as outlined in the introduction.  
However, combining these control strategies with strong interactions for the preparation and stabilization of correlated states of matter remains challenging, on the one hand, due to the above-mentioned Floquet heating \footnote{The mitigation of Floquet heating by increasing $\Omega$ is limited, as for too large $\Omega$ new heating channels open, given by transitions to  highly excited orbitals frozen out in the undriven system \cite{sunOptimalFrequencyWindow2020}.}, but also due to heating as a result of imperfect adiabatic state preparation when passing a phase transition. So far, only the Floquet control of the bosonic Mott transition for a short time 
\cite{
lignierDynamicalControlMatterWave2007,
zenesiniCoherentControlDressed2009}
as well as the preparation of baby fractional Chern insulator states of up to 3 particles \cite{wangRealizationFractionalQuantum2024,
leonardRealizationFractionalQuantum2023} have been reported. 

\section{Driven Bose-Hubbard chain}
In the following, we will describe a strategy for the dissipative preparation of many-body states  that appear as gapped ground states of the approximate Floquet Hamiltonian $\Ho_\text{eff}$. To illustrate our approach, we will first consider as a simple model the driven Bose-Hubbard chain described by the Hamiltonian
            \begin{align}\label{eq:DrivenBoseHubbard}
                \HS(t)
                =
                -J\sum_{\langle l'l\rangle} \rme^{\rmi\theta_{l'l}(t)}\hat{b}^{\dagger}_{l'}\hat{b}_{l}
                +\frac{U}{2}\sum_{l}\hat{n}_{l}(\hat{n}_{l}-1).
            \end{align}
Here, on-site interactions are characterized by the Hubbard parameter $U$ and nearest-neighbor tunneling by the amplitude $J>0$ as well as by the Peierls phases $\theta_{l'l}(t)=\alpha\sin(\Omega t)(l'-l)$, which describe a homogeneous time-periodic force with angular frequency $\Omega$ and amplitude $\alpha\hbar\Omega/d$, with lattice constant $d$. The operators $\hat{b}_l^\dagger$,  $\hat{b}_l$, and $\hat{n}_l = \hat{b}^\dagger_l \hat{b}_l$ are creation, annihilation and number operators for bosons at site $l$, respectively, and $\langle l'l\rangle$ denotes directed pairs of neighboring sites. This model describes, for instance, atoms in optical lattices, for which a thermal bath can be realized by the coupling to a cloud of atoms of a second species (see, e.g.\ \cite{schnellFloquetheatinginducedBoseCondensation2023}). 

With respect to the site-occupation Fock basis $\{|\{n_l\}\rangle\}$, the matrix elements of $\hat{H}_\mu$, as they appear in Eq.~(\ref{eq:Q}), take the form 
$\langle\{n'_l\}|\Ho_\mu|\{n_l\}\rangle
=\frac{U}{2}\sum_l \langle\{n'_l\}|\hat{n}_l(\hat{n}_l-1)|\{n_l\}\rangle - j_{s\mu}\sum_{\langle l'l\rangle}\langle\{n'_l\}|\hat{b}^{\dagger}_{l'}\hat{b}_{l}|\{n_l\}\rangle$,
 where  $s=+1$ ($-1$) for tunneling to the right (left) and $j_\mu\equiv J\mathcal{J}_\mu(\alpha)$ with Bessel function $\mathcal{J}_\mu$ \cite{eckardtColloquiumAtomicQuantum2017}. The time-averaged Hamiltonian 
    \begin{align}\label{eq:HF0}
        \Heff = \hat{H}_0 = - J_\text{eff} \sum_{\langle l'l\rangle} \hat{b}^{\dagger}_{l'} \hat{b}_{l}
        + \frac{U}{2} \sum_l \hat{n}_l (\hat{n}_l - 1),
    \end{align}
describes a Hubbard chain with modified tunneling parameter $J_\text{eff}=J \mathcal{J}_0(\alpha)$, whose absolute value drops from $J$ for $\alpha=0$ to $0$ for $\alpha\approx 2.4$. At integer filling and repulsive interactions, $U>0$, its ground state undergoes a quantum phase transition from a gapless superfluid to a gapped Mott-insulator, when $U/J_\text{eff}$ exceeds a critical parameter $u_c\approx 3.3$ \cite{ejimaDynamicPropertiesOnedimensional2011}. The Floquet control of this transition at large driving frequencies, $\hbar\Omega\gg U,J$, was proposed already more than two decades ago in Ref.~\cite{eckardtSuperfluidInsulatorTransitionPeriodically2005} and soon after observed experimentally \cite{zenesiniCoherentControlDressed2009}.

We focus on the deep Mott insulating regime at unit filling, $U \gg J_\text{eff}$, where the effective ground state can be approximated by a product state with one particle per site \cite{elstnerDynamicsThermodynamicsBoseHubbard1999}. Excited states contain at least one particle-hole excitation (PHE), created by moving a particle from one site to an already occupied site. One PHE costs an energy of about $U$, leading to an excitation gap. The energy of two PHEs is $2U$ if they are independent and $3U$ if both particle-excitations share the same site, and so on, giving rise to the band structure sketched in Fig.~\ref{fig:SpectrumSketch}. The widths of the bands result from the kinetic energy of delocalized particle and hole excitations. The resonant excitation of $n$ particle-hole pairs leads to the aforementioned avoided level crossings between states of different ``photon'' number $m$. For $\hbar\Omega\gg U$, this corresponds to an $n$-th order process with a small coupling matrix element $\sim J^n/(\hbar\Omega)^{n-1}$.

\section{Open system\label{sec:open_system}}
To stabilize the ground state against resonant heating into higher bands at the avoided crossings, the system shall be weakly coupled to a thermal environment, so that the total Hamiltonian reads
           $ 
                \hat{H}_\text{tot}(t)=\HS(t)+\HSB+\HB
           $, 
 where the bath is modeled as a collection of non-interacting harmonic oscillators $\HB=\sum_{l\alpha}\mathcal{E}_{\alpha} \hat{b}^{\dagger}_{l\alpha}\hat{b}_{l\alpha}$, with $\hat{b}^{\dagger}_{l\alpha}$ and $\hat{b}_{l\alpha}$ being the bosonic creation and annihilation operators at site $l$ and mode $\alpha$. The bath is in a thermal state $\rho_B\propto \exp(-\beta \Ho_B)$ with an inverse temperature $\beta$.  
The system-bath interaction is assumed to be given by $\HSB = \sum_{l}\hat{S}_l\otimes\hat{B}_l$, with system operators $\hat{S}_l$ and bath operators $\hat{B}_l=\sum_{\alpha}c_{\alpha} (\hat{b}^{\dagger}_{l\alpha} + \hat{b}_{l\alpha})$. For definiteness, we will consider $\hat{S}_l=\hat{n}_l$. Such a coupling to the local occupation number $\hat{n}_l$ is natural, for instance, in systems of ultracold atoms in optical lattices that are coupled via short-ranged collisions to the particles forming the thermal bath \cite{schnellFloquetheatinginducedBoseCondensation2023, schnellDissipativePreparationFloquet2024}. Note that this coupling operator remains invariant under transformations to (or from) a frame of reference, in which the forcing is not described by time-dependent Peierls phases $\theta_{l'l}(t)$, but rather by a time-dependent on-site potential.
We will, moreover, assume an Ohmic spectral function $ J(E) \equiv \sum_{\alpha}c_{\alpha}^2 \delta(E-\mathcal{E}_{\alpha})  = \gamma E\rme^{-|E|/E_{\mathrm{cut}}}$ with cutoff energy $E_\text{cut}$ (reflecting a finite spectral width of the bath) and dimensionless system-bath coupling strength $\gamma$.

In the weak coupling regime, the dynamics of the reduced system density matrix is governed by a Floquet-Born-Markov master equation \cite{blumelDynamicalLocalizationMicrowave1991,
kohlerFloquetMarkovianDescriptionParametrically1997,
grifoniDrivenQuantumTunneling1998,
breuerTheoryOpenQuantum2002,
honeStatisticalMechanicsFloquet2009}. Expressing the system's density operator in the Floquet basis, $\hat{\rho}(t)=\sum_{ij}| u_i (t)\rangle\rho_{ij}(t)\langle u_j (t)|$, it takes the form
            \begin{align}\label{eq:Redfield}
                (\diffd_t +\rmi\varepsilon_{ij}/\hbar)\rho_{ij}(t)
                =
                -\sum_{kn}
                &\left\{R_{inkn}\rho_{kj}(t) + R_{jknk}^{*}\rho_{in}(t)\right. \notag\\
                    -\rho_{kn}&(t)[\left. R_{njki}+R_{kinj}^{*}]\right\},
            \end{align}
where $\varepsilon_{ij}\equiv\varepsilon_i -\varepsilon_j$. The dissipative processes are covered by the super-operator matrix elements $R_{ijkn}=\sum_{l\mu} S^{ij}_{(l)}(\mu)S^{kn *}_{(l)}(\mu)G (\varepsilon_{nk}^{(-\mu)})$. Here, we defined the Fourier components of the system coupling operators $S^{nk}_{(l)}(\mu)=\frac{1}{T}\int_0^T \langle u_n(t)|\hat{S}_l|u_k(t) \rangle\rme^{-\rmi \mu \Omega t}\diffd t$ and the one-sided Fourier transform $G(E)=\int_0^\infty C(\tau) \rme^{-\rmi E\tau/\hbar}\diffd \tau$ of the bath correlation function $C(\tau)=\operatorname{Tr}_{\mathrm{B}}\{\hat{B}_{l}(\tau)\hat{B}_{l}\hat{\rho}_{\mathrm{B}}\}/\hbar^2$, with Heisenberg operators $\hat{B}_{l}(\tau)=\exp[\rmi\HB\tau/\hbar]\hat{B}_{l}\exp[-\rmi\HB\tau/\hbar]$. Moreover, $\varepsilon_{nk}^{(-\mu)}=\varepsilon_{nk}-\mu\hbar\Omega$ denote the energy change in the bath for a transition assisted by $\mu$ energy quanta from the drive. These equations were obtained employing the ``modified rotating wave approximation'': Assuming the driving frequency to be large compared to the system-bath coupling, the latter was averaged over one driving period \cite{honeStatisticalMechanicsFloquet2009}, making $R_{ijkn}$ time-independent and with that also $\rho_{ij}(t)$ in the long-time limit.

For even weaker system bath coupling, which is small compared to the quasienergy level splitting in the system, we can perform a rotating-wave (or secular) approximation \cite{honeStatisticalMechanicsFloquet2009}. In this limit, the off-diagonal matrix elements of the density operator decouple from the diagonal ones and decay. In the long-time limit, we, therefore, find states that are diagonal with respect to the exact Floquet states $|u_i(t)\rangle$, namely $\rho(t)=\sum_i P_i(t) |u_i(t)\rangle\langle u_i(t)|$, where we have introduced the probabilities $P_i=\rho_{ii}$ for finding the system in state $|u_i(t)\rangle$. They solve the \emph{Floquet-Pauli rate equation} \cite{Kohn2001}
\begin{equation}\label{eq:Pauli}
\dot{P}_i=\sum_j(R_{ij}P_j-R_{ji}P_i),
\end{equation}
which results from the Redifield equation within the  rotating-wave approximation (\ref{eq:Redfield}). Here, the rates are given by $R_{ij}=\sum_{\Delta m}R_{ij}^{\Delta m}$, where 
$
                R_{ij}^{\Delta m}=\sum_l
                2\pi g(\varepsilon_i-\varepsilon_j+\Delta m\hbar\Omega)|S^{ij}_{(l)}(
                \Delta m)|^2
 $
is the rate for a transition with energy change $\varepsilon_j-\varepsilon_i-\Delta m\hbar\Omega$ in the bath and $g(E)=J(E)n(E)$ depends on both the bath's spectral density $J(E)\propto \gamma$ and the temperature-dependent distribution function $n(E)=1/(\mathrm{e}^{\beta E}-1)$. The stationary solutions of the Floquet-Pauli Rate equation, obtained by setting $\dot{P}_i=0$, are independent of the coupling strength $\gamma$ (as every term on the right-hand-side of the equation is proportional to $\gamma$). Since also the rotating-wave approximation becomes exact in the limit $\gamma\to 0$, Eq.~(\ref{eq:Pauli}) accurately describes the stationary ($t\to\infty$) solution for $\gamma\to 0$.

Note that in a system of many interacting particles, the regime in which the Pauli rate equation (\ref{eq:Pauli}) would be valid, is neither realistic nor of interest: It is not realistic, as the quasienergy splitting at the tiny-avoided level crossings becomes extremely small.  Moreover, it is not desired, as a state that is diagonal with respect to the exact Floquet states $|u_i(t)\rangle$ would correspond to a high-energy state that is rather different from a desired low-energy state with a large occupation of the ground state of $\Ho_\text{eff}$. Namely, as a result of resonant coupling, the exact Floquet states will be superpositions of eigenstates of $\Ho_\text{eff}$ with very different energies.

\section{Dissipative suppression of Floquet heating}\label{sec:DissipativeSupressionFloquetHeating}
When two states of different ``photon'' number $m$ are resonantly coupled and hybridize at an avoided crossing of the quasienergy spectrum (see inset of Fig.~\ref{fig:SpectrumSketch} and discussion above) \footnote{In the large-system limit we might find that the effective ground state does not form individual avoided crossings with highly excited states, but rather couple to a continuum of such states. In this case the time scale of Floquet heating is not anymore determined by the coupling matrix element between two states alone, which is directly reflected in the width of the avoided crossing, but involves also the density of states like in Fermi's golden rule.}, we can distinguish two different regimes with respect to the coupling to the bath \cite{honeStatisticalMechanicsFloquet2009}: In the limit of ultra-weak system-bath coupling, where the bath-induced dynamics is slow compared to the time scale of the coherent coupling between the two levels (i.e.\ the time scale of Floquet heating), the density matrix will approach diagonal form with respect to the hybridized ``adiabatic'' basis given by the (actual) Floquet states (black dashed lines in the inset of Fig.~\ref{fig:SpectrumSketch}).
In the opposite regime, where the system-bath coupling is fast compared to the coherent coupling of the two states, it approaches diagonal form with respect to the ``diabatic'' basis (colored lines), given by states of definite ``photon'' number $m$ corresponding to the eigenstates of $\Ho_\text{eff}$. This effect relies on the fact that in the large-frequency regime of Floquet engineering, defined by the condition (\ref{eq:ConditionFlouqetEngineering}), the two states that form the avoided crossing are rather different, so that the coupling of these states via the coupling operator $\hat{S}_l$ becomes negligibly small \cite{honeStatisticalMechanicsFloquet2009}
\footnote{The fact that the system's density operator becomes diagonal with respect to the diabatic basis states \cite{honeStatisticalMechanicsFloquet2009}, resembles the quantum Zeno effect \cite{MisraSudarshan1977}. As the bath cannot couple the two states via the operators $\hat{S}$, it ''probes'' their population rather than their coherent superpositions.}.

The latter regime, where the coupling to the bath destroys coherences between eigenstates of $\Ho_\text{eff}$ which are resonantly coupled by the drive, corresponds to a situation where Floquet heating is suppressed by the coupling to the bath.
In Fig.~\ref{fig:FloquetSpace} this suppression is represented by purple crosses on top of the gray double arrows that stand for the coherent resonant excitation processes responsible for Floquet heating. In Ref.~\cite{shiraiEffectiveFloquetGibbs2016}, it was shown for the example of a spin chain that this effect can actually be used for stabilizing an effective Gibbs state with respect to $\Ho_\text{eff}$ at a temperature close to that of the environment (see also Ref.~\cite{moriFloquetStatesOpen2023} for a more general discussion). Here, in turn, we discuss a general strategy for achieving a large occupation of a gapped ground state of $\Ho_\text{eff}$. It works also in regimes where an approximate thermal occupation is difficult to achieve, as will be detailed below.

\section{Effective rate equation}
We will now investigate the properties of the bath that are required for approximately preparing the effective ground state $|\tilde{v}_0\rangle$ of $\Ho_\text{eff}$ or, more precisely, the corresponding approximate Floquet state $|v_0(t)\rangle=\Uo_F^{\mathrm{eff}}(t)|\tilde{v}_0\rangle$.
We will, at this point, assume that Floquet heating is suppressed by sufficiently strong system-bath coupling, as discussed above. In that case, we can set up an approximate Floquet-Born-Markov master equation like Eq.~(\ref{eq:Redfield}), but with respect to the approximate Floquet states $|v_\kappa(t)\rangle$ (rather than for the exact Floquet states $|u_n(t)\rangle$).
We, moreover, assume the effective energy gap $E_\text{gap}$ that separates the effective ground state from excited states of $\Ho_\mathrm{eff}$ to be large compared to the system-bath coupling, so that any coherences between the effective ground state and excited states decay in the long-time limit \cite{honeStatisticalMechanicsFloquet2009} and the system approaches a state
$\rho = P_0|v_0\rangle\langle v_0|+\sum_{\kappa,\lambda \ne 0} \eta_{\kappa,\lambda}|v_\kappa\rangle\langle v_\lambda|$, with probability $P_0$ of finding the system in the effective ground state. Since we are interested in the regime $1-P_0\ll1$ and do not require a precise description of the subspace of excited states, we, moreover, approximate $\eta_{\kappa,\lambda}\approx\delta_{\kappa\lambda}P_\kappa$. In the large-system limit, where excited states approach a continuum, this approximation does not provide an accurate description of the effective excited states, but it still serves us to estimate $P_0$. This assumption is confirmed when comparing the results to simulations of the full Floquet-Born-Markov equation (\ref{eq:Redfield}).

An equation for the probabilities $P_\kappa$ is obtained from a secular approximation with respect to all approximate Floquet states $|v_\kappa\rangle$, leading to the \emph{effective Pauli rate equation}
\cite{honeStatisticalMechanicsFloquet2009}
\begin{align}\label{eq:EffectivePauliEquation}
                \dot{P}_\kappa=
                \sum_\lambda (R_{\kappa \lambda}P_{\lambda}-R_{\lambda \kappa}P_{\kappa})=0,
            \end{align}
with rates $R_{\kappa\lambda}=\sum_{\Delta m}R_{\kappa\lambda}^{\Delta m}$ that are defined exactly like in the Floquet Pauli rate equation (\ref{eq:Pauli}), but now with the exact Floquet states $|u_i(t)\rangle$ replaced by the approximate Floquet states $|v_\kappa(t)\rangle$. That is,  
$ 
                R_{\kappa\lambda}^{\Delta m}=\sum_l
                2\pi g(E_\kappa-E_\lambda+\Delta m\hbar\Omega)|S^{\kappa\lambda}_{(l)}(
                \Delta m)|^2
 $ 
is the rate for a transition with energy change $E_\lambda-E_\kappa-\Delta m\hbar\Omega$ in the bath. Moreover, the matrix elements $S^{\kappa\lambda}_{(l)}(\Delta m)$ are defined as before, but now with respect to the $|v_\kappa(t)\rangle$. They become small in case the effective energy differences $|E_\kappa-E_\lambda|$ become large compared to the effective energy change that results from applying the coupling operator $\hat{S}_{l\mu}=\frac{1}{T}\int_0^T \Uo_F^{\text{eff}\dagger}\hat{S}_l\Uo_F^{\text{eff}}\rme^{-\rmi \mu\Omega t}\diffd t$ to $|\tilde{v}_\lambda\rangle$ or $|\tilde{v}_\kappa\rangle$. For transitions from or to the effective ground state, $|\tilde{v}_0\rangle$, this energy change corresponds to $\mathcal{E}_{\hat{S}_{l\mu}}=\mathcal{E}_{\hat{S}_{\mu}}$ as defined in Eq.~(\ref{eq:EnergyScale}), where we dropped the subscript $l$ as this energy scale should not depend noticeably on the site index $l$.
Since all terms in Eq.~(\ref{eq:EffectivePauliEquation}) are proportional to the system-bath coupling strength $\gamma$, its solutions are independent of it. 
Note that Eq.~(\ref{eq:EffectivePauliEquation}) is used only to estimate $P_0$ in the regime where $P_0$ is large.

\section{Floquet-space resolved rate equation}

An intuitive picture for the processes with $\Delta m\ne0$ is to associate them with rates between states $|v_\kappa m\rrangle$ of different ``photon'' number $m$ in Floquet space. Namely, postulating the rate equation 
\begin{equation}\label{eq:RateEquationPhotons}
\dot{P}_\kappa^{m}=\sum_{\lambda\Delta m} [R_{\kappa\lambda}^{\Delta m}P_\lambda^{m-\Delta m}-R_{\lambda\kappa}^{\Delta m}P_\kappa^m]
\end{equation}
for the corresponding probabilities $P_\kappa^m\in[0,1]$.  We recover equation (\ref{eq:EffectivePauliEquation}) by setting 
\begin{equation}\label{eq:Pkm}
P_\kappa\equiv\sum_m P_\kappa^m.
\end{equation}
This picture resembles the Floquet picture of the two-time formalism, where the coherent evolution of a system is obtained from the dynamics generated by the quasienergy operator $\hat{Q}$ in Floquet space \cite{BreuerHolthaus1989}. For each state $\kappa$ the initial probabilities $P_\kappa^m$ can be chosen arbitrarily, as long all $P_\kappa^m$ are positive and obey Eq.~(\ref{eq:Pkm}). For instance, one can assume the system to be completely in the zero-photon sector, $P_\kappa^{m=0}=P_\kappa$ and $P_\kappa^{m\ne0}=0$ for $t=0$, or, alternatively, distribute the probabilities equally over all ``photon'' numbers,  $P_\kappa^{m}=P_\kappa/(2M+1)$ with $m=-M,-M+1,\ldots M$ and $M\to\infty$. Note that the intuitive picture introduced here for the effective Pauli rate equation can equally be applied to the Floquet Pauli rate equation (\ref{eq:Pauli}).

The advantage of thinking in terms of the ``photon''-number resolved rate equation (\ref{eq:RateEquationPhotons}) is that it resembles an equilibrium rate equation for states $(\kappa,m)$ with energy $E_{\kappa m}=E_\kappa+m\hbar\Omega$. Namely, for a transition $(\kappa,m)\to(\kappa',m')$ the associated energy change in the bath is given by $\Delta E_\mathrm{bath}= -(E_{\kappa' m'}-E_{\kappa m})$ and, as a consequence, the forward and backward rate obey the condition $R_{\kappa \kappa'}^{m-m'}/R_{\kappa' \kappa}^{m'-m}=\exp[-\beta(E_{\kappa m}-E_{\kappa' m'})]$. This allows one to use (part of) our intuition for equilibrium systems.  Despite these similarities, however, no Gibbs state is approached for the $P_{\kappa}^m$, since the ``energies'' $E_{\kappa m}$ are not bounded from below.

\section{Bath engineering}
Let us now formulate a general strategy for cooling a driven system into the gapped ground state of $\Ho_\text{eff}$. On the basis of the effective  Pauli rate equation (\ref{eq:EffectivePauliEquation}), we need to suppress transition rates $R_{\kappa 0}=\sum_{\Delta m}R_{\kappa 0}^{\Delta m}$ from the effective ground state to excited states relative to rates $R_{0\kappa}=\sum_{\Delta m}R_{0\kappa}^{\Delta m}$ for processes populating the ground state. The former can be viewed as bath-assisted (or incoherent) Floquet heating, whereas the latter describe cooling. On the right-hand side of Fig.~\ref{fig:FloquetSpace} the most relevant processes of the former type are represented by black curved (single) arrows, labeled~(b-d), while the latter processes are indicated by the thick black curved single arrows labeled~(a).

Apart from fine tuning or selection rules, there are three generic possibilities for suppressing a rate $R_{\kappa\lambda}^{\Delta m}$:
\begin{itemize}
\item[(i)] The bath energy change is large compared to the bath's spectral width, 
\begin{equation}
    |E_\lambda-E_\kappa-\Delta m\hbar\Omega|\gg E_\mathrm{cut}.
\end{equation}
\item[(ii)] The bath energy change is negative and large compared to the bath temperature, 
\begin{equation}
    0 < -(E_\lambda-E_\kappa-\Delta m \hbar\Omega)\gg \beta^{-1}.
\end{equation} 
\item[(iii)] The transition matrix elements $S^{\kappa\lambda}_{(l)}(\Delta m)$ are small because the effective energy change is large compared to the typical energy changes induced by the system-bath coupling, 
\begin{equation}
    |E_\kappa-E_\lambda|\gg\mathcal{E}_{\hat{S}_{\Delta m}}.
\end{equation}
\end{itemize}

In order to achieve a large ground-state probability $P_0$, we first require the bath temperature to be small compared to the energy gap, $\beta^{-1}\ll E_\mathrm{gap}$. In this way all transitions from the ground state to excited states with $\Delta m\ge 0$ [corresponding to processes (d) in Fig.~\ref{fig:FloquetSpace}] are thermally suppressed by condition (ii). We must also suppress transitions out of the ground state to excited states at lower photon number, i.e.\ with $\Delta m<0$. In the considered high-frequency regime, we can assume $\hbar\Omega\gg \mathcal{E}_{\hat{S}_{\Delta m}}$. By requiring the frequency to be large also compared to the spectral width of the bath, $\hbar\Omega\gg E_\mathrm{cut}$, transitions to states at lower photon number are either suppressed by (i) or (iii): Namely, transitions to highly excited states with effective energies $E_\kappa-E_0 \sim \hbar\Omega$ [described by processes (c) in Fig.~\ref{fig:FloquetSpace}], which are not necessarily suppressed by $E_\mathrm{cut}$ via (i), are suppressed by small transition matrix elements via (iii) (This is also a prerequisite for the suppression of coherent Floquet heating as described in Sec.~\ref{sec:DissipativeSupressionFloquetHeating}). Transitions to lower lying excited states with $E_\kappa-E_0\lesssim\mathcal{E}_{\hat{S}_{\Delta m}}$ [corresponding to processes (b) in Fig.~\ref{fig:FloquetSpace}] and matrix elements that are not suppressed via (iii) are, in turn, suppressed by the bath cutoff via (i). At the same time, in order not to suppress transitions into the ground state via (i) and (iii), we also require $E_\mathrm{gap}\lesssim \mathcal{E}_{\hat{S}_{\Delta m}}, E_\mathrm{cut}$. All in all, these conditions stemming from the effective Pauli rate equation (\ref{eq:EffectivePauliEquation}) can be summarized as 
\begin{align}\label{eq:RateConditions}
\beta^{-1}\ll E_\mathrm{gap}\lesssim \big\{\mathcal{E}_{\hat{S}_{\Delta m}} , E_\mathrm{cut} \big\} \ll \hbar\Omega. 
\end{align}
These conditions are rather natural, as large frequencies are already required for the condition (\ref{eq:ConditionFlouqetEngineering}) for Floquet engineering and  $\mathcal{E}_{\hat{S}_{\Delta m}}$ will correspond to a typical intensive energy scale of $\Ho_0$, as the operators $\hat{S}_l$ usually describe single-particle processes. Finally, it is also natural to require the bath temperature $\beta^{-1}$ to be small compared to the many-body energy gap $E_\text{gap}$.

In addition to the conditions (\ref{eq:ConditionFlouqetEngineering}) and (\ref{eq:RateConditions}), we also have to require the system-bath-coupling strength $\gamma$ to obey
\begin{align}\label{eq:Coupling}
\gamma_\mathrm{heating}\ll\gamma \ll \gamma_\mathrm{gap},
\end{align}
where for $\gamma=\gamma_\mathrm{heating}$ and $\gamma=\gamma_\mathrm{gap}$ the system-bath coupling would be comparable to the energy scale of Floquet heating and the energy gap, respectively. Together with the requirement $\hbar\Omega\gg \mathcal{E}_{\hat{S}_{\Delta m}}$, which is included already in the relations (\ref{eq:RateConditions}) and which ensures negligible bath-induced coupling between resonantly coupled states, this condition ensures that (coherent) Floquet heating is suppressed, as discussed in Sec~\ref{sec:DissipativeSupressionFloquetHeating}.

It is important to note that conditions (\ref{eq:RateConditions}) and (\ref{eq:Coupling}) are much less restrictive than those required for preparing an approximate Gibbs state of $\Ho_\mathrm{eff}$ in our system, following the strategy of Refs.~\cite{shiraiEffectiveFloquetGibbs2016, moriFloquetStatesOpen2023}. First, the system bath coupling is only needed to be small compared to the energy gap $E_\mathrm{gap}$, but not with respect to the level splittings between excited states of $\Ho_\mathrm{eff}$, which become exponentially small with increasing system size. Second, a thermal state would require that zero-``photon'' transitions out of the ground state [which for the interesting regime of small temperatures $\beta^{-1}\ll E_\text{gap}$ are suppressed exponentially via (ii)] are dominant compared to non-zero-``photon'' transitions [which are suppressed by either (i), (ii) or (iii)], $R_{\kappa0}^0\gg \sum_{\Delta m\ne 0}R_{\kappa0}^{\Delta m}$.

\section{Application to driven Bose-Hubbard chain}
The proposed strategy summarized in Eqs.~(\ref{eq:ConditionFlouqetEngineering}), (\ref{eq:RateConditions}) and (\ref{eq:Coupling}) does not rely on the details of the system (except from the assumption of $\Ho_\text{eff}$ possessing a many-body energy gap). Let us now test it for the concrete example of the driven open Bose-Hubbard chain introduced above. 
The effective Hamiltonian is gapped in the Mott regime, for integer filling and $U/J_\text{eff}>u_c$, with an energy gap $E_\text{gap}\le U$ associated with the cost of creating a particle-hole excitation. Here also $\mathcal{E}_{\hat{S}_{\Delta m}}$ is on the order of $U$, since the largest energy increase associated with applying the Fourier components of the coupling operators, $\frac{1}{T}\int \, \hat{U}_F^{\text{eff}\dag}\hat{S_l}\hat{U}_F^\text{eff}\rme^{-\rmi\Delta m \Omega t}\diffd t$, to the effective ground state $|\tilde{v}_0\rangle$ is the creation of particle-hole excitations. Thus, together with the already assumed high-frequency condition for Floquet engineering, $\hbar\Omega\gg U,J$, we have already $E_\mathrm{gap}\lesssim \mathcal{E}_{\hat{S}_{\Delta m}}\ll \hbar\Omega$ and it remains to choose the bath parameters $E_\mathrm{cut}$, $\beta$, and $\gamma$, so that $E_\mathrm{gap}\lesssim E_\mathrm{cut} \ll\hbar\Omega$, $\beta^{-1}\ll E_\mathrm{gap}$ as well as $\gamma_\mathrm{heating}\ll\gamma \ll \gamma_\mathrm{gap}$. Below we choose $\hbar\Omega=25J$, $\beta^{-1}=0.3J$, and $E_{\mathrm{cut}} = U$ (except for $U<J$, where we set $E_{\mathrm{cut}} = J$), and study the behavior as a function of $\gamma$.

    \begin{center}
        \begin{figure}
            \includegraphics[width=\linewidth]{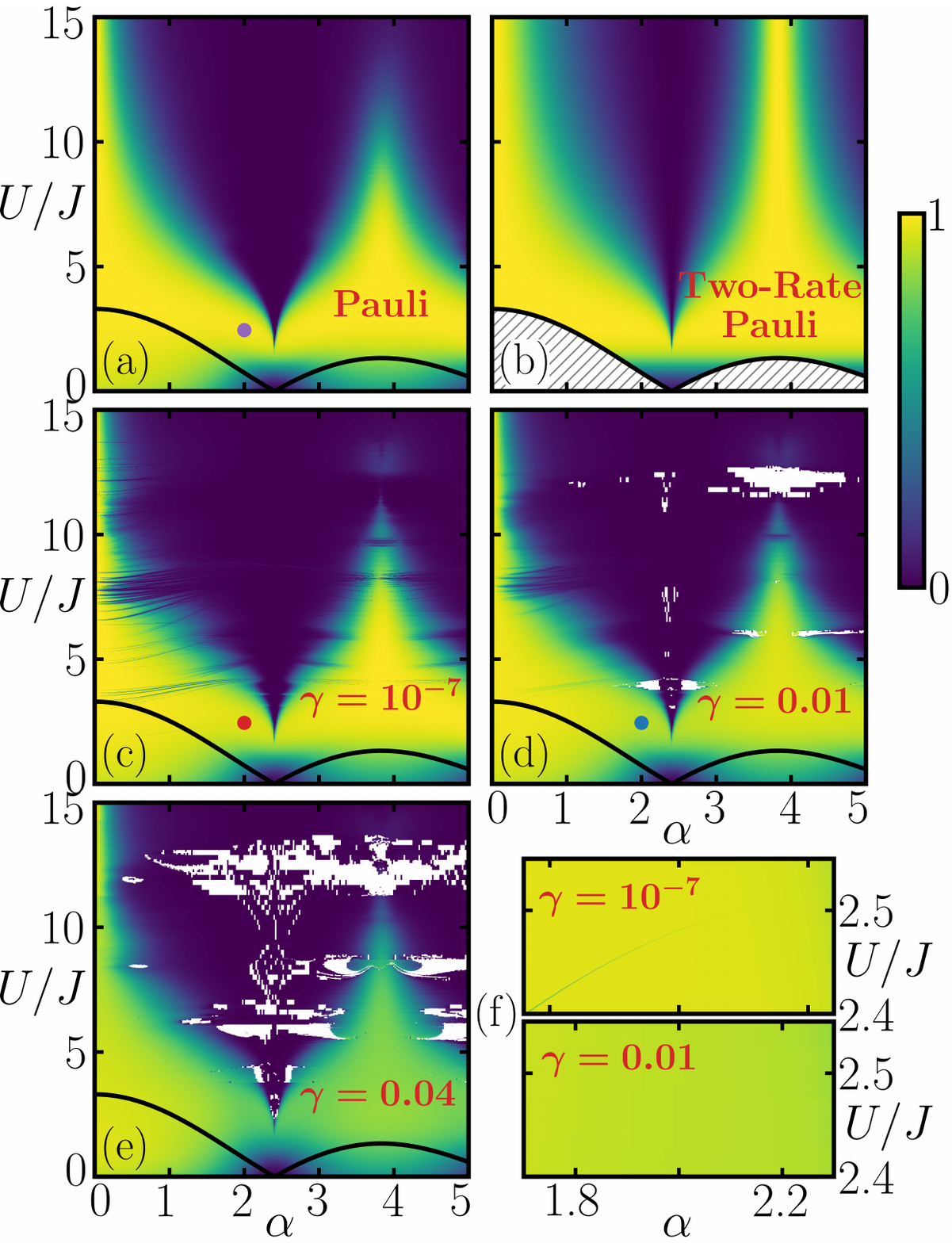}
            \caption{\label{fig:DoubleMapL=5} Steady-state ground-state probability $P_0=\langle v_0|\hat{\rho}|v_0\rangle$ for 5 particles on 5 sites, with $\hbar\Omega=25J$, $\beta^{-1}=0.3J$, $E_{\mathrm{cut}} = \mathrm{max}(U,J)$,  calculated using (a) the Pauli equation (\ref{eq:EffectivePauliEquation}), (b) a simplified two-state rate equation, (c,d,e) the full Redfield equation (\ref{eq:Redfield}) at coupling strengths $\gamma=10^{-7}$, $\gamma=0.01$, and $\gamma=0.04$, (f) a zoom on the full Redfield results in the regime marked by the colored dots in (c,d) for $\gamma=10^{-7}$ and $\gamma=0.01$. The black line indicates the superfluid to Mott-insulator transition. The white areas in (d,e) mark regions where the Redfield equation gives non-positive density operators.}
        \end{figure}
    \end{center}

In Fig.~\ref{fig:DoubleMapL=5} we plot the steady-state ground-state probability $P_0=\langle v_0|\hat{\rho}|v_0\rangle$ for a small system of five particles on five sites as a function of the dimensionless strengths of interactions, $U/J$, and driving, $\alpha$. Above the black line, the ground state of $\Ho_\mathrm{eff}$ is a Mott insulator (in the thermodynamic limit). Panel (a) is obtained from the simple Pauli rate equation (\ref{eq:EffectivePauliEquation}) for the approximate Floquet states, whereas (c,d,e) result from the full Redfield equation (\ref{eq:Redfield}) for three different coupling strengths, $\gamma=10^{-7}$, $\gamma = 10^{-2}$, and $\gamma=4\cdot 10^{-2}$, respectively. While the rate equation shows very good agreement with the solutions of the full equation for most of the parameters, one can note pronounced differences for the weaker coupling $\gamma$ in the form of horizontal lines of strongly reduced ground state population in panels (c) and (d). These lines follow resonances and reflect Floquet heating,  which is not captured by the effective rate equation. However, we can see that for the stronger  coupling, $\gamma=0.04$, Floquet heating is indeed strongly suppressed by the coupling to the bath, as these lines are hardly visible anymore in panel (e). This regime is to good approximation captured by the results obtained using the effective rate equation shown in panel (a). This confirms that the strategy proposed above indeed works.

\begin{center}
         \begin{figure}
             \includegraphics[width=\linewidth]{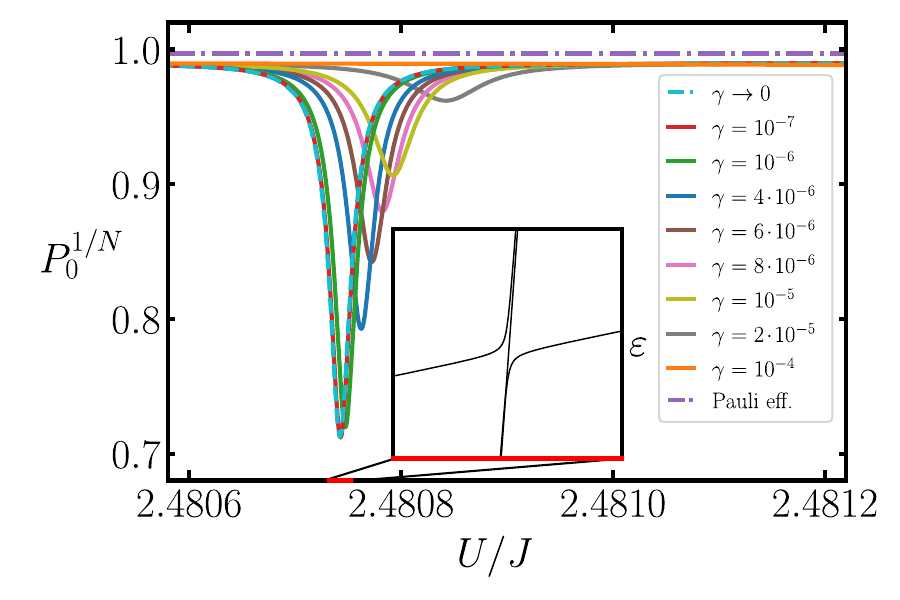}
\caption{\label{fig:UScan} 
Steady-state ground-state probability $P_0=\langle v_0|\hat{\rho}|v_0\rangle$ for the Bose-Hubbard model with 5 particles on 5 sites for $\hbar\Omega=25 J$, $\beta^{-1}=0.3J$, $\alpha = 2$ as a function of $U/J$ near the resonance marked by the colored dots in Fig.~\ref{fig:DoubleMapL=5}. The inset shows part of the quasienergy spectrum, where the effective ground state forms an avoided crossing with an effective excited state with five particles on one site. The hybridization of both states causes the dip in the population $P_0$ for small~$\gamma$. }
 \end{figure}
     \end{center}

Let us now have a closer look at the suppression of a particular resonance. In an experiment, one would choose parameters, where Floquet heating is already weak. Let us, therefore, focus on the parameter regime marked by the colored dots in panels (a,c,d,e) in Fig.~\ref{fig:DoubleMapL=5}. Here, for $\alpha=2$, tunneling is already markedly suppressed by the driving, so that the system is deep in the Mott regime already for $U/J\approx 2.5$. For such small values of $U/\hbar\Omega\approx 0.1$, we only find a weak resonance, corresponding to the excitation of states with 5 particles on a single lattice sites, with interaction energy $\approx 10 U\approx \hbar\Omega$. In panel (f) we zoom into panels (c,d). While the resonance is clearly visible for $\gamma = 10^{-7}$, it is suppressed for $\gamma = 10^{-2}$. In Fig.~\ref{fig:UScan}, we show $P_0$ in the steady state close to the resonance as a function of $U/J$. The solid curves labeled by finite $\gamma$ are obtained from the full Redfield equation and confirm the suppression of Floquet heating in the steady state, visible as an increase of $P_0$, with increasing $\gamma$. We also plot results obtained from the Floquet-Pauli rate equation (\ref{eq:Pauli}) (cyan dashed line), capturing the limit $\gamma\to0$, where Floquet heating is not suppressed. It agrees well with the results for $\gamma=10^{-7}$.  Moreover, the purple dash-dotted line is computed from the effective rate equation (\ref{eq:EffectivePauliEquation}), build on the assumption that Floquet heating is suppressed. It agrees rather well with the results for stronger system bath coupling $\gamma=10^{-4}$.

\begin{center}
         \begin{figure}
             \includegraphics[width=\linewidth]{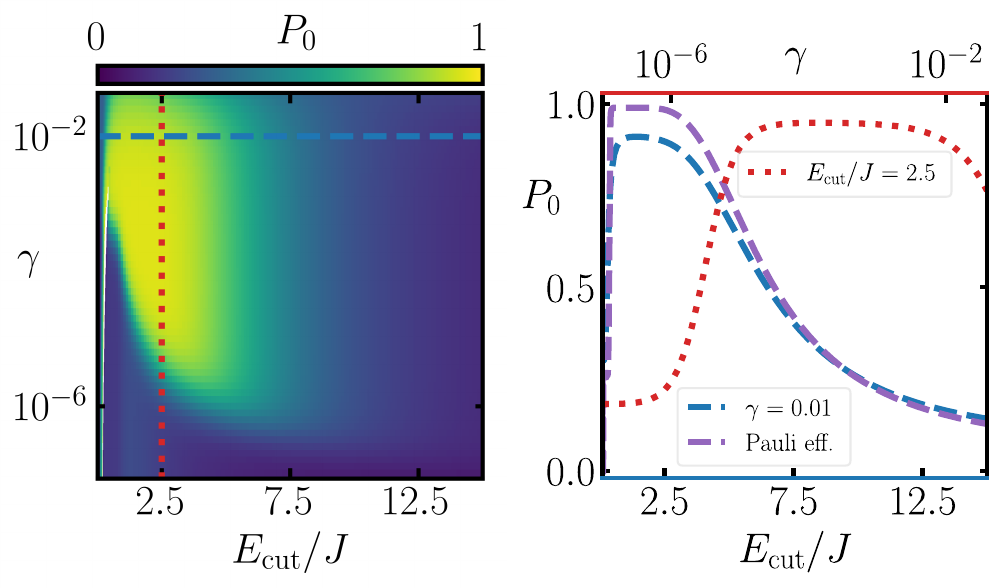}
\caption{\label{fig:Robustness} 
Steady-state ground-state probability $P_0=\langle v_0|\hat{\rho}|v_0\rangle$ for the Bose-Hubbard model with 5 particles on 5 sites for $\hbar\Omega=25 J$, $\beta^{-1}=0.3J$, $\alpha = 2$ and $U=2.480743J$ (resonance condition, marked by colored dots in Fig.~\ref{fig:DoubleMapL=5}). Left: $P_0$ as function of coupling strength $\gamma$ and cutoff energy $E_\text{cut}$. Right: Blue dashed and red dotted lines show a horizontal and a vertical cut indicated in the left figure by the same line style. The purple dashed line follows the horizontal cut, but is obtained from the effective Pauli rate equation. It agrees well with the full Redfield result.}
 \end{figure}
     \end{center}

The fact that for the simulations shown in Figs.~\ref{fig:DoubleMapL=5} and \ref{fig:UScan} we have confirmed the targeted large steady-state occupation of the effective ground state for sufficiently strong system-bath coupling $\gamma$, is a consequence of the fact that we chose the parameters according to the general conditions (\ref{eq:ConditionFlouqetEngineering}),  (\ref{eq:RateConditions}), and (\ref{eq:Coupling}). This can be confirmed by varying not only the coupling strength $\gamma$, but also the cutoff energy $E_\text{cut}$ controlling the bath's spectral width, as it is done in Fig.~\ref{fig:Robustness} for parameter values close to the resonance investigated in Fig.~\ref{fig:UScan}. In the left panel, as well as from the horizontal cut through it shown by the blue dashed line in the right panel, we can see that large target state populations $P_0$ are found in a broad window of cutoff energies $E_\text{cut}$. However, as expected $P_0$ drops both for too small cutoff energies, since a spectral width that is small compared to the energy gap hinders cooling processes bridging the energy gap [type (a) in Fig.~\ref{fig:FloquetSpace} (right)], as well as for too large cutoff energies, where bath-induced heating processes [type (b) in Fig.~\ref{fig:FloquetSpace} (right)] become stronger. These effects are well described also by the effective rate equation (purple dashed lines in the right panel of Fig.~\ref{fig:Robustness}). The right panel of Fig.~\ref{fig:Robustness} also shows a vertical cut through the left panel, where the coupling strength $\gamma$ is varied for fixed $E_\text{cut}$. Again, we find a broad window where heating is efficiently suppressed. However, this vertical cut overestimates the suppression of Floquet heating. Namely, we can see in Fig.~\ref{fig:UScan} that the resonance is shifted when  $\gamma$ is varied over several orders of magnitude, which is a Lamb-shift effect. As a result $P_0$ reaches its plateau value a bit earlier as it would with the shift. The height of the plateau is practically not modified, since for $\gamma\gtrsim10^{-4}$ the dip in $P_0$ has essentially vanished. This is also the reason, why for the horizontal cut, where $E_\text{cut}$ is varied at large $\gamma=10^{-2}$, the plateau is not modified by Lamb-shift effects.

If Floquet heating is suppressed by a sufficiently large $\gamma$, we have seen that the ground-state population is fairly well described by the effective rate equation (\ref{eq:EffectivePauliEquation}). Having a closer look at the this equation (see Appendix \ref{sec:TwoRate}  for details), one can see that the dominant transitions into the effective ground state are \emph{zero}-``photon'' ($\Delta m=0$) processes from states with a single PHE [labeled $(a)$ in Figs.~\ref{fig:SpectrumSketch} and \ref{fig:FloquetSpace} (right)], whereas the dominant processes out of the ground state are \emph{single}-``photon'' ($\Delta m=-1$) transitions to states with a single PHE  [labeled $(b)$ in Figs.~\ref{fig:SpectrumSketch} and \ref{fig:FloquetSpace} (right)]. The results from this minimal two-rate model, plotted in Fig.~\ref{fig:DoubleMapL=5}(b), agree rather well with those for the full rate equation shown in Fig.~\ref{fig:DoubleMapL=5}(a). The occupation $P_0$ is, thus, mainly determined by the competition of a zero- and a single-``photon'' process. As a consequence, the population of the eigenstates of $\Ho_\mathrm{eff}$ do not correspond to an approximate Gibbs state at the inverse bath temperature $\beta$. Moreover, as a result of the $\Delta m=-1$ processes, maintaining the steady state causes a steady energy flow from the drive, through the system, into the bath, which we can interpret as Floquet heating of the bath. This emphasizes the non-equilibrium nature of the process used here to approximately stabilize the effective ground state.

\section{Disspative preparation of Floquet fractional Chern insulator}

As a second example, let us now consider the preparation of a bosonic fractional Chern insulator in a square lattice with Floquet engineered homogeneous magnetic flux. That is, the effective Hamiltonian shall correspond to the Harper-Hofstadter Bose-Hubbard model.  We consider the driving scheme of Ref.~\cite{aidelsburgerMeasuringChernNumber2015}. It is again described by the driven Bose-Hubbard model (\ref{eq:DrivenBoseHubbard}), but now with the site indices $l=(m,n)$ with the $x$ and $y$ position given by $m=1,2,\ldots L_x$ and $n=1,2,\ldots L_y$, respectively, 
describing a square lattice of rectangular shape with open boundary conditions. 
The drive is given by a static superlattice of amplitude $\Delta$ and a lattice potential of amplitude $\kappa$ moving in diagonal direction \cite{aidelsburgerMeasuringChernNumber2015} so that the lattice is modulated resonantly with frequency $\Omega=\Delta/\hbar$ and site-dependent phase lag. It is described by the time-dependent Peierls phases 
$\theta^x_{mn}(t)=
    -\frac{\sqrt{2}\kappa}{\hbar\Omega}
    \big\{    
        \sin\left(\frac{m\pi}{2}\right)\sin(\Omega t+g(n))
        -\cos\left(\frac{m\pi}{2}\right)\cos(\Omega t-g(n))
    \big\}
+\Omega t(-1)^{m+1}$
and 
$\theta^y_{mn}(t)=
    -\frac{\sqrt{2}\kappa}{\hbar\Omega}
    \big\{
        \sin\left(\frac{m\pi}{2}+\frac{\pi}{4}\right)\sin\left(\Omega t+g(n)+\frac{\pi}{4}\right)
        -\cos\left(\frac{m\pi}{2}+\frac{\pi}{4}\right)\cos\left(\Omega t-g(n)-\frac{\pi}{4}\right)
    \big\}$
for tunneling from site $l=(m,n)$ in positive $x$ and $y$ direction to site $l'=(m+1,n)$ and $l'=(m,n+1)$,   respectively, with $g(n)=\pi/4-n\pi/2$.

The effective Hamiltonian obtained from time averaging is then given by the Harper-Hofstadter Bose-Hubbard Hamiltonian,
\begin{align}\label{eq:Heff_HHH}
    \Heff
    =&
    -\sum_{m,n}
    \left(
        J_x\rme^{\rmi n\Phi} \hat{a}^\dagger_{m+1,n}\hat{a}_{m,n} 
        +J_y\hat{a}^\dagger_{m,n+1}\hat{a}_{m,n}
        +\mathrm{h.c.}
    \right)
    \notag
    \\&
    +\frac{U}{2}\sum_{m,n}\hat{n}_{m,n}(\hat{n}_{m,n}-1),
\end{align}
with $J_x=J\mathcal{J}_1\left(\sqrt{2}\kappa/\hbar\Omega\right)$, $J_y=J\mathcal{J}_0\left(\sqrt{2}\kappa/\hbar\Omega\right)$ and $\Phi=\pi/2$. The Peierls phases describe a  homogeneous magnetic flux per plaquette corresponding to a quarter of the flux quantum $2\pi$. 
We will assume $\kappa=\hbar\Omega$, giving rise to approximately isotropic tunneling amplitudes, $J_x\approx J_y\approx 0.55J\equiv J_\text{eff}$. 
For strong interactions, the effective Hamiltonian possesses a gapped ground state given by a bosonic fractional Chern insulator at a filling factor of $\nu=1/2$ particles per magnetic unit cell (given by $2\pi/\Phi=4$ plaquettes) \cite{Hafezi2007}. This state has been realized experimentally in small systems of two particles in an optical lattice \cite{leonardRealizationFractionalQuantum2023} and a superconducting circuit \cite{wangRealizationFractionalQuantum2024}. Since for larger system sizes adiabatic preparation (passing a finite-size precursor of a topological phase transition) rapidly becomes challenging (and also Floquet heating more relevant), so far only fractional Chern insulator states of at most three particles \cite{kwanPfaffianQuantumHall2026} have been realized in Floquet engineered optical-lattice systems.
The dissipative state preparation scheme described here provides an alternative preparation scheme that is not limited by the system size.

In the following, we consider a system of $N=3$ particles on $L_x\times L_y = 7\times 5$ lattice sites, i.e.\ on $6\times4=24$ plaquettes or $24 /4 = 6$ magnetic unit cells for quarter flux, $\Phi/2\pi=1/4$, corresponding to half filling, $\nu=1/2$. Our dissipative preparation scheme is not limited to such small systems, however, this is the largest size for which we are able to treat the open Floquet system numerically. 

\begin{center}
    \begin{figure}
        \includegraphics[width=1.0\linewidth]{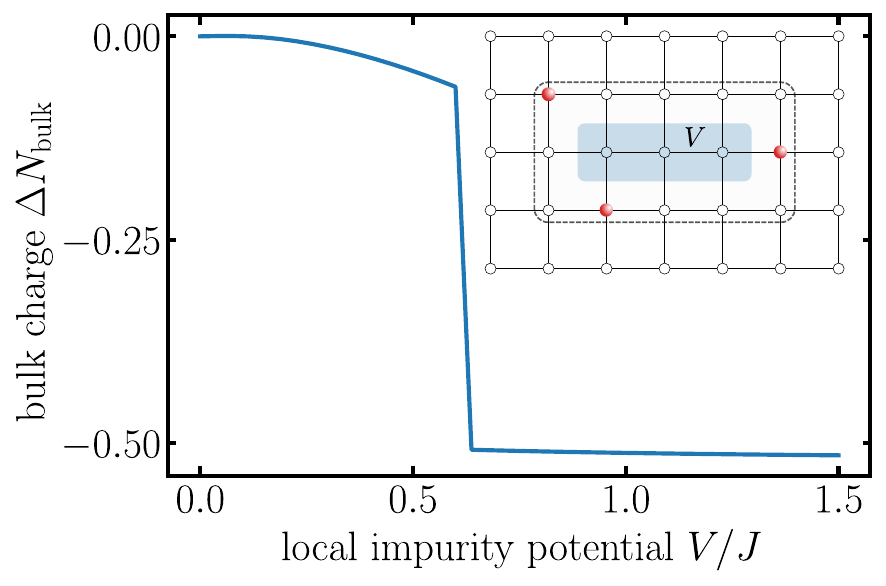}
\caption{\label{fig:V_pinning} $\Delta N_\text{bulk}$ vs.\ $V/J$. Inset: Sketch of the system of $N=3$ particles (symbolized by the red bullets) on $L_x\times L_y=7\times 5$ lattice sites. On the three central sites (blue shaded area) a repulsive potential $V$ is applied and the resulting change of particle number inside the dashed line, $\Delta N_\text{bulk}$, recorded. Parameters: $J_x=J_y=0.55J$, $U=6.777J$.}
    \end{figure}
\end{center}

Before investigating the preparation numerically, let us first briefly verify that for the system parameters chosen, the ground state of the effective Hamiltonian $\Ho_\text{eff}$ indeed shows signatures expected for a $\nu=1/2$ Laughlin-like bosonic fractional quantum Hall state. For that purpose, we study the response to a local potential $V$ applied to the central three sites of the lattice inside the blue shaded area in the inset of Fig.~\ref{fig:V_pinning} \cite{Raciunas2018,Wang2022, wangRealizationFractionalQuantum2024}. In Fig.~\ref{fig:V_pinning} we plot how the particle number in the vicinity of this perturbation, defined by the sites inside the the dashed line in the inset of the figure, changes with $V/J$ relative to its value for $V=0$. The response, denoted $\Delta N_\text{bulk}$, remains small, up to a value of $V/J\approx 0.6$, where it suddenly jumps to $\approx -1/2$, to remain there when $V$ is increased further. While the initial stiffness against the perturbation is a signature for a bulk energy gap, the drop can be interpreted with the creation of a quasi-hole excitation associated with a ``charge'' of $-\nu=-0.5$, as it is expected for the targeted $\nu=1/2$ fractional quantum Hall state. (Note that $V/J\approx 0.6$ does not directly reflect the bulk gap, as it depends also on the energy needed to place a quasi-particle of opposite ``charge'' $+1/2$ on the edge of the system). 

Let us now apply the bath-engineering strategy developed above to the preparation of the Floquet engineered FCI. We consider the same coupling to the bath as before for the Bose-Hubbard chain. We choose $\kappa=\hbar\Omega$ and $\Omega = 25 J$ for the amplitude and the frequency of the drive and vary $U/J$ near values of $U/J=6.8$, where the effective FCI-type ground state is resonantly coupled to a band of excited states. We choose this value on purpose to demonstrate that for weak coupling to the bath, the system will suffer from Floquet heating. In an experiment with such a small system, one would naturally rather choose parameters, where such resonances are avoided. However, with increasing system size, it will become more and more difficult to completely avoid resonances, as the number of states (as well as the density of states) increases exponentially. Therefore, we deliberately choose parameters for which Floquet heating is present.

\begin{center}
 \begin{figure}
            \includegraphics[width=\linewidth]{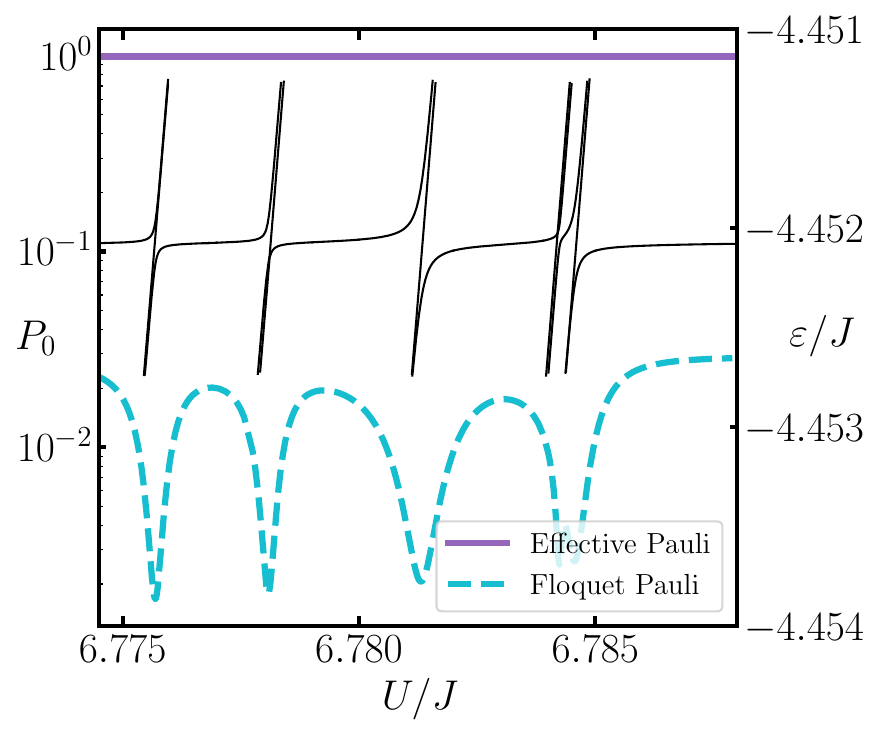}
  \caption{\label{fig:FCI_Uscan} 
  Population of the effective fractional Chern-insulator ground state, $P_0=\langle v_0|\hat\rho|v_0\rangle$,  computed vs.\ $U/J$ for the open driven Bose-Hubbard model used to Floquet engineer the Harper-Hofstadter Bose-Hubbard model at quarter flux (colored lines). The results obtained from the Floquet-Pauli equation are valid in the limit $\gamma\to0$, whereas the effective Pauli equation approximately describes larger values of $\gamma$, for which Floquet-heating is suppressed. The thin black lines show a horizontal slice of the quasienergy spectrum centered around the effective ground state energy. Parameters: $N=3$, $L_x\times L_y=7\times 5$, $\hbar\Omega=25J$, $\beta^{-1}=0.01J$, $E_\mathrm{cut}=2.0J$, $\kappa=\hbar\Omega$.}
 \end{figure}
\end{center}

For the chosen parameters, the energy gap of the effective Hamiltonian is given by $E_\text{gap}\approx 0.06J\approx 0.1 J_\text{eff}$. To fulfill the first requirement of the hierarchy of energy scales (\ref{eq:RateConditions}), we choose the temperature well below the gap, $\beta^{-1}=0.01J$. The energy scale of the bath coupling operator $\mathcal{E}_{\hat{S}}$ is that of the single-particle energy changes in the system, which are of the order $U$ or $J$, so that it is equivalent to the requirement for the high-frequency expansion, $\hbar\Omega\gg U, J$, which is fulfilled by the chosen frequency $\hbar\Omega = 25 J$. This also ensures $E_\text{gap}\ll\hbar\Omega$. Finally, we choose $E_\text{cut}=2J$, i.e. larger than the energy gap but still small compared to $\hbar\Omega$. 

Since solving the full Redfield equation is not feasible for the Hilbert-space size, we solve both the Floquet Pauli equation, which describes the open system in the limit of $\gamma\to0$ and the effective Pauli rate equation which describes the non-equilibrium populations of the eigenstates of $\Ho_\text{eff}$ for values of $\gamma$ that are strong enough to suppress Floquet heating. For both approaches we plot the population of the effective ground state, $P_0=\langle v_0|\hat{\rho}|v_0\rangle$, in Fig.~\ref{fig:FCI_Uscan} as a function of the interaction strength $U/J$. We also include a slice of the quasienergy spectrum, showing the effective ground state as well as effective excited states of lower photon number (a larger part of the quasienergy spectrum is given in Fig.~\ref{fig:SpecFCI} in Appendix \ref{sec:FullSpec}). The avoided crossings, where the effective ground state hybridizes with excited states, correspond to Floquet heating. This is clearly visible as dips in $P_0$ at the same positions for $\gamma\to0$.  However, the population remains small also between these dips. For stronger $\gamma$ the dips will eventually disappear, this regime is described by the effective Pauli rate equation (see for instance Fig.~\ref{fig:Robustness} (right), where we compared the effective rate equation to the full Redfield equation for the Bose-Hubbard chain) for which we were able to compare the effective rate equation to the the full Redfield equation). In this regime, we find large probabilities $P_0$ for finding the system in the  effective fractional Chern insulator state, which shows that our approach allows to stabilize this state.

\begin{center}
 \begin{figure}
            \includegraphics[width=0.8\linewidth]{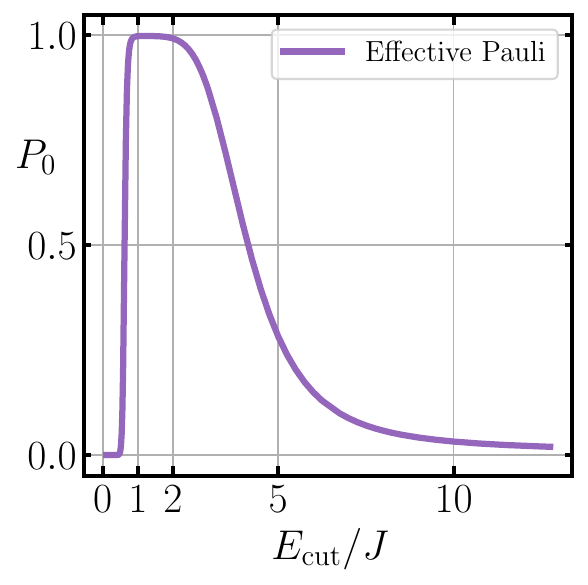}
  \caption{\label{fig:FCI_Ecut_scan}Steady-state ground-state probability as a function of the bath cutoff $E_\mathrm{cut}$, calculated using the effective Pauli equation. The interaction is fixed to $U/J=6.777$; all other parameters are as in Fig.~\ref{fig:FCI_Uscan}.}
 \end{figure}
\end{center}

Note that it is not guaranteed that the effective rate equation provides large ground-state occupations $P_0$.  Namely, except for Floquet heating within the system (which will be suppressed for sufficiently large $\gamma$), the effective Pauli rate equation still describes the non-equilibrium nature of the driven-dissipative system by capturing all $\mu$-``photon'' processes involving system and bath. The fact that we find large values of $P_0$ is rather a consequence of having chosen the parameters according to the general strategy worked out above, i.e.\ according to the hierarchy (\ref{eq:RateConditions}) of energy scales. In order to test that, in Fig.~\ref{fig:FCI_Ecut_scan} we plot the ground-state population $P_0$ as a function of the bath's cutoff energy $E_\text{cut}$. We fix $U/J=6.777$ (close to the second resonance in Fig~\ref{fig:FCI_Uscan}), while all other parameters correspond to those of Fig.~\ref{fig:FCI_Uscan}. It becomes apparent that large values of $P_0$ are, like for the open driven Bose-Hubbard chain [Fig.~\ref{fig:Robustness} (right)], found only in a window of cutoff energies that is consistent with the conditions (\ref{eq:RateConditions}). 

\section{Conclusion and outlook}
We have proposed a general strategy for the dissipative preparation and stabilization of gapped ground states of Floquet engineered effective Hamiltonians, as they are described by low-order high-frequency expansions. It is based on the coupling to a thermal bath and relevant for current experiments in various quantum-simulation platforms based on neutral atoms or photons. Here, Floquet engineering is an essential tool for the implementation of system properties, such as the presence of artificial magnetic fields for charge-neutral particles, that naturally are not present in these systems and otherwise hard to achieve. Our approach provides a solution to two relevant problems that severely limit the preparation of correlated states by means of Floquet engineering in interacting systems: on the one hand it suppresses Floquet heating and on the other it guides the system into the target state without the need of adiabatic state preparation, which becomes challenging beyond system sizes of a few particles only. In turn, the dissipative scheme proposed here is scalable with system size. Moreover, it requires a hierarchy of energy scales that is based on the separation of three energy scales, given by temperature, energy gap, and driving frequency, in ascending order, which is rather natural and, therefore, does not pose a major limitation. 

We have demonstrated our scheme for two examples. The first one is the preparation of a bosonic Mott insulator in a one-dimensional Bose-Hubbard chain. For this model, we were able to verify it by solving the full Floquet-Born-Markov master equation. Moreover, we showed that two simple rate equations allow to describe the system in two relevant parameter regimes: In the limit of ultraweak system-bath coupling, where Floquet heating spoils the stabilization of the target state, the system is described by the Floquet-Pauli rate equation. In turn the effective Pauli rate equation provides a description of the non-equilibrium state of the system at larger coupling strength, where Floquet heating is suppressed. With the help of these equations we were then able to show that also a dissipative preparation of a Floquet engineered bosonic fractional Chern-insulator state is possible and even facilitated by a smaller many-body energy gap. 

In this work, we have assumed the coupling to a thermal bath with energy cutoff. Such environments can be realized and engineered in systems of neutral atoms via the coupling of the system to a bath given by a second atomic species (whose spectral properties can be shaped with the help of lattice potentials giving rise to finite-width Bloch bands, see, e.g., Ref.~\cite{schnellDissipativePreparationFloquet2024}).  However, in other platforms, it might be easier to consider engineered environments, as they are more naturally implemented in superconducting circuits, e.g.\ via the dispersive coupling to driven-dissipative cavities \cite{Zhang2019, Zhang2022, PetiziolEckardt2022}. For such a scenario a strategy for preparing Floquet engineered fractional Chern insulators via reservoir engineering was recently proposed in Ref.~\cite{steinfadtDissipationassistedPreparationFloquetLaughlin2026}. There, the suppression of Floquet heating was not discussed, assuming it can be avoided by tuning away from resonances or for small systems with the frequency large compared to the system's total spectral band width. The authors only address the suppression of Floquet heating by referring to the first arXiv-version of the present paper. Investigating, whether Floquet heating can indeed be suppressed also by such engineered reservoirs in the same way as by the thermal reservoirs discussed here, is an interesting question for future research. 

The simulations presented here were restricted to small systems. In fact, the simulation of open many-body quantum systems is generally difficult. On the one hand, an accurate (microscopic) description of the bath is required, as far from equilibrium the details of the environment beyond its temperature are required. However, the standard approaches for deriving equations of motion for the open system, such as (Floquet)-Born-Markov theory, require the computation of all energy eigen (or Floquet) states of the system. Moreover, also the integration of the equations of motion is a difficult many-body problem. Thus, both for deriving and solving equations of motion, one will typically encounter excited and non-equilibrium many-body states, which can be expected to feature large entanglement. This renders the application of tensor-network ansätze inefficient, at least during the transient approach of the target state. On the other hand, the theoretical analysis presented here, does not rely on any assumption about the system size. Therefore, the scheme should equally work also for large systems. While the numerical verification is, as just argued, extremely challenging, this problem can readily be addressed experimentally in engineered quantum many-body systems. Such an experiment would correspond to a genuine quantum simulation.

\begin{acknowledgments}
    We thank Francesco Petiziol, Alexander Schnell, and Luis Steinfadt for discussions. This work was supported by the Deutsche Forschungsgemeinschaft (DFG, German Research Foundation) via the Research Unit FOR 5688 (Project No.\ 521530974) and by Berlin Quantum, an initiative endowed by the Innovation Promotion Fund of the city of Berlin.
\end{acknowledgments}

\section*{Data availability}

The data that support the findings of this study are available from the corresponding author upon reasonable request. 

\bibliography{library}

\onecolumngrid
\vspace{1cm}
\twocolumngrid

\appendix

\section{Quasienergy spectrum}
\label{sec:FullSpec}
\begin{center}
        \begin{figure*}
        \includegraphics[width=0.8\linewidth]{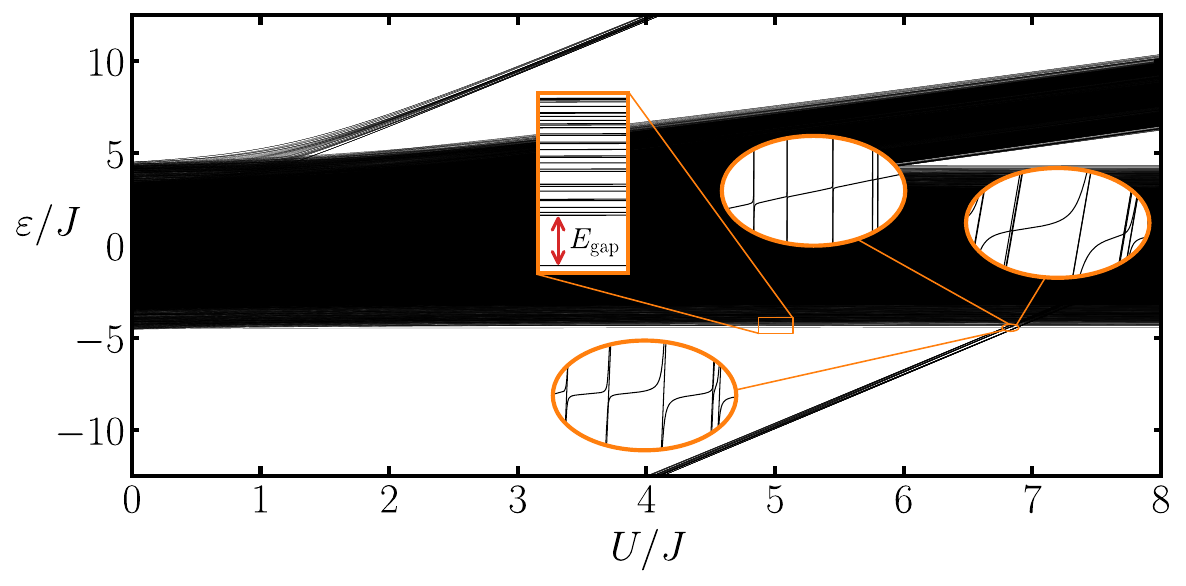}
  \caption{\label{fig:SpecFCI} Quasienergy spectrum of the driven Bose-Hubbard model used to Floquet engineer the Harper-Hofstadter Bose-Hubbard model. The parameters are identical to those of Fig.~\ref{fig:FCI_Uscan}. The most left inset shows the many-body gap that opens at large interactions, $U/J\gg 1$. The other insets show avoided crossings of the ground state with excited states. The inset in the bottom mid corresponds to the avoided crossings shown in Fig.~\ref{fig:FCI_Uscan}.}
 \end{figure*}
    \end{center}

Figure~\ref{fig:SpecFCI} shows the quasienergy spectrum  of the driven Bose-Hubbard model used to Floquet engineer the Harper-Hofstadter Bose-Hubbard model.

\section{Two-rate model}
\label{sec:TwoRate}

In the Mott regime, $U/J_\mathrm{eff}>u_c$, the rates $R_{\kappa\lambda}^{\Delta m}$ decay exponentially with the number of (``photon''-assisted) tunneling processes required to connect two states. Therefore, the dynamics involving the effective ground state are dominated by two main transitions, labeled $(a)$ and $(b)$ in Fig.~\ref{fig:SpectrumSketch}. The $a$-process connects the ground state to states with a single PHE via a single ``photon''-free ($\Delta m=0$) tunneling event. The $b$-process connects the ground state to the first excited band in the photon subspace below ($\Delta m= -1$) via a ``photon''-assisted hop. At low temperatures, $
\beta^{-1}\ll E_\mathrm{gap}$, $a$ is unidirectional toward the ground state and constitutes the dominant cooling channel, and $b$ is unidirectional toward the excited state and represents the main heating channel. As all the different photon subspaces are equivalent, the $a$ and $b$ transitions form a closed loop. 
In the following we estimate  the approximate coupling matrix elements $S_{(l)}^{\kappa\lambda}(\mu)$ for these two processes.

Deep in the Mott regime at unit filling, we can treat the tunnel kinetics as a perturbation, so that in the leading zeroth order the ground state of $\Ho_\mathrm{eff}$ is a product state with one particle per site,
        \begin{align}
            |\tilde{v}_\mathrm{G}\rangle^{(0)} = |1,1,\dots,1\rangle.
        \end{align}
Zeroth order (unperturbed) excited states are given by states with PHE excitations on top of this state. States with a single PHE,
        \begin{align}
            |\tilde{v}_{l_1,l_2}\rangle^{(0)} = \frac{1}{\sqrt{2}} \hat{a}_{l_1}^\dagger \hat{a}_{l_2} |\tilde{v}_\mathrm{G}\rangle^{(0)},
        \end{align}
with particle and hole sitting next to each other, i.e.\ $|l_1-l_2|=1$, are coupled to the ground state by a single tunneling process. The coupling to other unperturbed states is of higher order. The above mentioned processes $(a)$ and $(b)$ correspond to such leading-order transitions to states with a single PHE.

In order to compute the transition rates
\begin{equation}
R_{\kappa\lambda}^{\Delta m}=\sum_l
                2\pi g(E_\kappa-E_\lambda+\Delta m\hbar\Omega)|S^{\kappa\lambda}_{(l)}(
                \Delta m)|^2
\end{equation}
for the processes $(a)$ and $(b)$, with $g(E)=J(E)n(E)$, with thermal occupation function $n(E)=1/(\rme^{\beta E}-1)$ and spectral density $J(E)=\gamma E \rme^{-|E|/E_\mathrm{cut}}$ of the bath, we have to evaluate the matrix elements
\begin{equation}
S^{ \kappa \lambda}_{(l)}(\mu)=\frac{1}{T}\int_0^T \langle v_\kappa(t)|\hat{n}_l|v_\lambda(t) \rangle\rme^{-\rmi \mu \Omega t}\diffd t
\end{equation}
for the approximate Floquet states $|v_\mathrm{G}(t)\rangle^{(\nu)}  =  \Uo_F^\mathrm{eff}(t)| \tilde{v}_\mathrm{G}\rangle^{(\nu)}$ and $ |v_{l_1,l_2}(t)\rangle^{(\nu)} = \Uo_F^\mathrm{eff}(t) |\tilde{v}_{l_1,l_2}\rangle^{(\nu)}$.

At large driving frequencies, $\hbar\Omega\gg \Ho_\mu$ we can expand the exponential in the effective micromotion operator (\ref{eq:HF}) to obtain
    \begin{align}
    \Uo_F^\mathrm{eff}(t)\simeq 1 -\sum_{\mu\neq 0} \frac{\rme^{\rmi \mu\Omega t}}{\mu\hbar\Omega} \Ho_\mu,
    \end{align}
where $\Ho_\mu=-\sum_{\langle l'l\rangle}j_{s\mu}\hat{b}_{l'}^\dagger \hat{b}_l$,  $s=+1~(-1)$ for tunneling to the right (left) and $j_\mu=J\mathcal{J}_\mu(\alpha)$. 
This yields the coupling matrix elements of the on-site density operator between the approximate Floquet modes 
        \begin{align}\label{eq:supp:FourierMatrixElement}   
            {}^{(0)}\!\langle {v}_{l_1,l_2}(t)| 
             \hat{n}_i |{v}_\mathrm{G}(t)\rangle^{(0)}
            \approx
            \sqrt{2}
            \sum_{\mu\neq 0} 
            \frac{j_{s\mu}}{\mu\hbar\Omega}
            \rme^{\rmi \mu\Omega t} 
            (\delta_{il_1}-\delta_{il_2}).
        \end{align}
We can read of the Fourier components and identify the full coupling matrix elements for the photon-assisted tunneling process $(b)$ in Fig.~\ref{fig:SpectrumSketch},
        \begin{align}
            S^{(b)} \sim  \sqrt{2}\frac{J\mathcal{J}_{1}(\alpha)}{\hbar\Omega}. 
        \end{align}

The corresponding matrix element for the zero-``photon'' process $(a)$, vanishes between the unperturbed states. Therefore, we take into account also the leading correction with respect to the tunneling term of the effective Hamiltonian Eq.~(\ref{eq:HF0}),  
        \begin{align}
            |\tilde{v}_{l_1,l_2}\rangle^{(1)} 
            &= |\tilde{v}_{l_1,l_2}\rangle^{(0)}
            - \sqrt{2}\frac{J\mathcal{J}_0(\alpha)}{U}\, |\tilde{v}_\mathrm{G}\rangle^{(0)},\\
            |\tilde{v}_\mathrm{G}\rangle^{(1)} 
            &= |\tilde{v}_\mathrm{G}\rangle^{(0)}
            + \sqrt{2}\frac{J\mathcal{J}_0(\alpha)}{U}\, |\tilde{v}_{l_1,l_2}\rangle^{(0)},
        \end{align}
where we have neglected admixtures of states to $|\tilde{v}_\mathrm{G}\rangle^{(1)}$, with two PHE as well as with a single PHE excitation but $|l_1-l_2|\ge2$, as they are irrelevant for the calculation of $ S^{(a)}$.
Taking into account only zeroth order micromotion $\hat{U}_F^\mathrm{eff} \simeq 1$, we find in leading order in $J$,
        \begin{align}
            {}^{(1)}\!\langle {v}_{l_1,l_2}(t)| \hat{n}_i |{v}_\mathrm{G}(t)\rangle^{(1)}
            \approx \sqrt{2}\frac{J\mathcal{J}_0(\alpha)}{U}(\delta_{il_1}-\delta_{il_2}),
        \end{align}
yielding the matrix element for process $(a)$,
        \begin{align}
            S^{(a)} \sim \sqrt{2}\frac{J\mathcal{J}_{0}(\alpha)}{U}.
        \end{align} 
   
To estimate the ground-state population, we retain only the $a$- and $b$-processes with transition energies $\Delta E_a\sim U$ and $\Delta E_b\sim \hbar\Omega-U$. 
This gives the estimate for the cooling rate, defined as the total rate for transitions from those excited state with a single PHE with $|l_1-l_2|=1$ to the ground state 
\begin{align}
R_\mathrm{cool} &= g(-U)\,|S^{(a)}|^2 + g(\hbar\Omega-U)\,|S^{(b)}|^2
\end{align}
as well as the total heating rate for reverse transitions,
\begin{align}
R_\mathrm{heat} &= g(U)\,|S^{(a)}|^2 + g(U-\hbar\Omega)\,|S^{(b)}|^2.
\end{align}

For the true eigenstates of $\Ho_\mathrm{eff}$ in the first excited band of a single PHE, both particle and hole are delocalized as a result of tunneling. They can, roughly, be thought of as coherent superpositions of states $|\tilde{v}_{l_1,l_2}\rangle^{(0)}$ with arbitrary $|l_1-l_2|$. Their total number is $N_\mathrm{ex}=L(L-1)$, for a system of $L$ sites. Assuming the fraction of states with $|l_1-l_2|=1$ that couple to the ground state, the number of which is $N'_\mathrm{ex}=2(L-1)$, to be roughly equal for all eigenstates of the first band, we expect  rates $R_\mathrm{cool}N'_\mathrm{ex}/N_\mathrm{ex}$ and $R_\mathrm{heat}N'_\mathrm{ex}/N_\mathrm{ex}$ for each of the true eigenstates. Taking into account only processes $(a)$ and $(b)$, the population of each excited state is then determined by the ground state population and the ratio of heating and cooling rate, so that the factors $N'_\mathrm{ex}/N_\mathrm{ex}$ drop out, 
\begin{align}
P_\mathrm{ex}=\frac{R_\mathrm{heat}}{R_\mathrm{cool}}P_0.
\end{align}
Together with the normalization condition $P_0+N_\mathrm{ex}P_\mathrm{ex}=1$, we finally obtain
\begin{align}\label{eq:EstimatedGroundStatePopulation}
P_0=\frac{1}{1+N_\mathrm{ex}\,R_{\mathrm{heat}}/R_{\mathrm{cool}}}\,.
\end{align}
This is the result plotted in Fig.~\ref{fig:DoubleMapL=5}(b). It provides rather good qualitative agreement with the more accurate results shown in the other panels.  

\end{document}